\documentstyle[aas2pp4]{article}

\def\B{{\bf B}}

\def\Gammad{\Gamma_{\rm d}}

\def\Gammas{\Gamma_{\rm s}}
\def \gta {\mathrel{\vcenter
     {\hbox{$>$}\nointerlineskip\hbox{$\sim$}}}}
\def\'#1{\ifx#1i{\accent"13\i}\else{\accent"13#1}\fi}

\def \lta {\mathrel{\vcenter
     {\hbox{$<$}\nointerlineskip\hbox{$\sim$}}}}
\def\ni{\noindent}
\def\o{{\rm o}}

\def\rhoic{\rho_{\rm ic}}
\def\to{t_{\rm o}}
\def\u{{\bf u}}
\def\VS{V\'azquez-Semadeni}

\def \m{\ifmmode {\rm M}_\odot\else M$_\odot$\fi}

\slugcomment{Submitted to {\it The Astrophysical Journal}}

\begin{document}

\title{On the Density Probability Function of Galactic Gas. I. Numerical
Simulations and the Significance of the Polytropic Index}

\author{John Scalo$^1$, Enrique V\'azquez-Semadeni$^2$, David Chappell$^1$, and
Thierry Passot$^3$}
\affil{$^1$Astronomy Department, The University of Texas at Austin}
\affil{e-mail: {\tt parrot,dwch@astro.as.utexas.edu}}
\affil{$^2$Instituto de Astronom\'\i a, UNAM, Apdo. Postal 70-264, M\'exico, D.\
F.\ 04510, M\'exico}
\affil{e-mail: {\tt enro@astroscu.unam.mx}}
\affil{$^3$Observatoire de la C\^ote d'Azur, B.P.\ 4229, 06304, Nice
             Cedex 4, France}
\affil{e-mail: {\tt passot@obs-nice.fr}}
Submitted to {\it The Astrophysical Journal}.

\begin{abstract}

We investigate the form of the one-point probability distribution function (pdf)
for the density field of the interstellar medium using numerical simulations
that successively reduce the number of physical processes included. 
Two-dimensional simulations of self-gravitating supersonic MHD turbulence, of
supersonic self-gravitating hydrodynamic turbulence, and of decaying Burgers
turbulence, produce in all cases filamentary density structures and evidence
for a power-law density pdf with logarithmic slope around $-1.7$. This suggests
that the functional form of the pdf and the general filamentary morphology are
the signature of the nonlinear advection operator. 

These results do not support
previous claims that the pdf is lognormal. A series of 1D simulations of forced
supersonic polytropic turbulence is used to resolve the discrepancy. They
suggest that the pdf is lognormal only for effective polytropic indices $\gamma=
1$ (or nearly lognormal for $\gamma\not=1$ if the Mach number is sufficiently
small), while power laws develop for densities larger than the mean if $\gamma<
1$. We evaluate the polytropic index for conditions relevant to the cool
interstellar medium using published cooling functions and different heating
sources, finding that a lognormal pdf may occur at densities between 10$^3$ and
at least 10$^4$ cm$^{-3}$. 

Several applications are examined. First, we question
a recent derivation of the IMF from the density pdf by Padoan, Nordlund \& Jones
because a) the pdf does not contain spatial information, and b) their derivation
produces the most massive stars in the {\it voids} of the density distribution.
Second, we illustrate how a distribution of ambient densities can alter the
predicted form of the size distribution of expanding shells. Finally, a brief
comparison is made with the density pdfs found in cosmological simulations.

\end{abstract}

%

\section{Introduction}
In order for stars or groups of stars to form, densities must be
sufficiently large, 
and so the statistical properties of the density field in the
interstellar medium must be 
intimately connected with the star formation behavior of galaxies.  
The density field is  coupled in a nonlinear manner 
to the velocity distribution of the gas, so the statistics of the
density field can 
potentially serve as a diagnostic between different physical processes
that might play a role 
in controlling the dynamics of interstellar material.  For example,
magnetically-dominated 
turbulence might exhibit significantly different density statistics
than non-magnetic 
turbulence.  Thus the density distribution could shed light on basic
unsolved questions 
concerning star-forming regions, such as the nature and maintainance
of highly supersonic 
motions observed at all but the smallest scales.  The form of the
density statistics might 
also be important  in controlling the ability of galactic gas to
spontaneously develop 
hierarchical structure, as suggested by \VS\ (1994). Padoan, Jones \&
Nordlund (1997) have 
shown how observations of extinction fluctuations by Lada et
al.\ (1994) can be used to 
constrain the statistics of the density field, while Padoan (1995) and
Padoan, Nordlund \& 
Jones (1997) have attempted to relate the density distribution
function to the stellar initial 
mass function, assuming a specific model for star formation.
Understanding the physics behind 
the one-point density distribution function is also important in a
cosmological context, since 
the form of this function in the nonlinear regime may be sensitive to
the initial fluctuation statistics (see sec. 4.4. below).
 
In the present paper we focus on the behavior of the one-point
probability distribution 
of gas densities in numerical simulations relevant to galactic gas.
This function, denoted 
$f(\rho)$, is defined here such that $f(\rho)d\rho$ measures the
fractional volume 
occupied by gas in the density range ($\rho,\ \rho+d\rho$), so
$f(\rho)$ is a probability 
density function (pdf) with units 1/density.  If the density pdf is
parameterized in some 
range of density as a power law with index $\theta$,
$f(\rho)\sim\rho^{\theta}$, then the 
fractional volume of space at density $\rho$ per unit {\it logarithmic}
density interval and the 
fractional {\it mass} of material per unit density interval would be a power
law with index $(\theta + 1)$.
 
\VS\ (1994) presented evidence from
two-dimensional numerical simulations of randomly-forced
turbulence that the density pdf may be lognormal
in form; i.e. that the
{\it logarithm} of the density has a probability density that is normal
(Gaussian).  However, his 
simulations were purely hydrodynamic (i.e., neglected self-gravity,
the magnetic field and 
the Coriolis force), isothermal, and were restricted to relatively small Mach
numbers compared to what is 
observed in most interstellar regions. \VS, Passot \& Pouquet (1995) 
have presented simulations incorporating all of these
physical effects, but discussed the density pdf only in passing.
Nordlund \& Padoan (quoted in Padoan et al. 1997a,b,c) have also found a 
lognormal density pdf in randomly forced isothermal three-dimensional 
simulations. These simulations include the magnetic 
field, and reach higher Mach  numbers, 
but contain no other astrophysically relevant physical ingredients.  
 
Part of the purpose of the present work is to examine whether
this lognormal density pdf 
persists in simulations that include most of the major physical
processes expected to play 
some role in galactic turbulence, including self-gravity, stellar
heating, the magnetic 
field, rotation, and explicit heating and cooling functions in the
energy equation, and which 
are carried out at large Mach numbers.  If the results 
are different, we want to find out why, and try to isolate the
dominant physical effect(s). In fact, Gotoh \& Kraichnan (1993) have
presented numerical evidence (and an interpretation in terms of the
``mapping'' closure) that the pdf for the Burgers equation (which is
pressureless) is a power-law at high densities, rather than a lognormal.
 
In section 2.1 we first show that simulations of the most
complex system do not result 
in a lognormal pdf, but instead show evidence for power law behavior
at densities above the 
mean.  In order to understand the reasons for the different result and
to isolate the 
physical process(es) that are most responsible for this behavior, we
next successively reduce 
the system by removing physical processes.  In sec. 2.2 we examine
simulations similar to the 
original set, but without magnetic fields and rotation.  Again we find
the power law pdf, 
with a similar slope and density range as before.  As a more radical
reduction of the system, 
in sec. 2.3 we study the density pdf for simulations that contain no
physics at all except for 
nonlinear advection in the continuity and momentum equations (Chappell
\& Scalo 1997). 
These simulations contain no pressure, and so are equivalent to a gas
with effective polytropic index equal to zero.\footnote{The polytropic
  index characterizes the 
  response of the gas to quasi-static compression or expansion in
  thermal equilibrium, and is of course 
  different from the ratio of specific heats which is, for example, 5/3
  for a monotonic gas 
  with no internal degrees of freedom.} Surprisingly, the
spatial density 
fields and the density pdfs resemble the earlier simulations at
densities above the mean, 
suggesting  that advection is the dominant process. The importance of
the advection term is well 
known in incompressible turbulence, in which even the pressure term can be 
incorporated into the nonlinear operator.
 
In order to understand these results, in section 3 we present
a few conclusions that can be drawn from a large 
study of one-dimensional forced turbulence by Passot \&
\VS\ (1997), who 
have investigated the statistics of the density and velocity 
fields for a wide 
range of Mach numbers and polytropic indices.  It turns out 
that a lognormal density 
pdf should only obtain at small Mach numbers or, for any value of the
Mach number, 
when the polytropic index is equal to unity. This explains the
discrepancy between 
the present simulations (all of which have small effective polytropic
indices) and previous 
work (which assumed $\gamma=1$).  We also discuss the expected value of the
polytropic index as a function of density and 
temperature, based on published cooling functions and various heating
sources and conclude that at densities less than about 10$^3$
cm$^{-3}$ the polytropic index should be 
significantly less than unity, favoring power-law density pdfs.  For
higher densities we show 
that saturation of the cooling function should lead to
$\gamma\approx1$ and therefore possibly a 
lognormal pdf.  However at still larger densities $\gamma$
gas-grain collisional cooling or heating 
should dominate, forcing $\gamma$ away from unity again.  
Some implications of these results are discussed in section
\ref{applications}. In particular, we argue that the distribution of masses of 
collapsing structures or stars (the IMF) cannot be derived from the
one-point density pdf, and illustrate how a distribution of densities might
affect the predicted size distribution of expanding shells driven by young
stars.

\section{Three types of simulations and their pdfs} \label{models}

\subsection{Self-gravitating magnetohydrodynamic \break\hfill 
models with rotation}   \label{MHDmod}

A first class of models considered in the present paper consists of the fully 
nonlinear self-gravitating magnetohydrodynamic (MHD) equations with added 
model source terms for radiative cooling, stellar and diffuse heating, 
and the Coriolis force, as described in Passot, \VS\ \& Pouquet (1995). 
The equations of this model, to which we will refer as the MHD model, are:

\begin{equation}
{\partial\rho\over\partial t} + {\nabla}\cdot (\rho\u) = \mu
\nabla^2 \rho, \label{MHDcont}
\end{equation}
\begin{eqnarray}
{\partial\u\over\partial t} + \u\cdot\nabla\u =
 -{\nabla P\over \rho} 
  - \Bigl({J \over M_a}\Bigr)^2 \nabla \phi +\nonumber \\
  {1 \over \rho}
  \bigl(\nabla \times {\bf B}\bigr) \times {\bf B}
  - 2 \Omega \times \u - {\nu_8} {\nabla^8\u}+\nonumber \\
  \nu_2 (\nabla^2 \u + \frac{1}{3} \nabla \nabla \cdot \u) \label{MHDmom} 
\end{eqnarray}
\begin{eqnarray}
{\partial e\over\partial t} + \u\cdot\nabla e = -(\gamma -1)
e\nabla\cdot \u + {\kappa}_T {\nabla^2e\over \rho} + \nonumber\\
({1 \over \rho})(\Gammad + \Gammas - \Lambda), \label{MHDener}
\end{eqnarray}
\begin{equation}
{\partial {\bf B}\over\partial t} = \nabla \times (\u \times {\bf B})
- {\nu_8} {\nabla^8{\bf B}} + \eta \nabla^2 \B, \label{MHDmagn}
\end{equation}
\begin{equation}
\nabla^2 \phi=\rho -1, \label{MHDpoisson}
\end{equation}
\begin{equation}
P=(\gamma-1)\rho e, \label{MHDeqstat}
\end{equation}
\begin{equation}
\Gammad({\bf x},t)={\Gamma_{\o}(\rho/\rhoic)^{a}}, \label{MHDgammad}
\end{equation}
\begin{equation}
\Gammas({\bf x},t) = 
\cases{\Gamma_1 \rho   & if $\rho({\bf x},t_0) >
        \rho_{\rm cr}$ \cr
        & and $0 < t-t_0 < \Delta t_s$\cr
        0               & otherwise\cr}, \label{MHDgammas}
\end{equation}
with
\begin{equation}
\Lambda= \rho^2 L(T), \nonumber
\end{equation}
and
\begin{equation}
L(T) = L_i T^{b_i} \ \ \ {\rm for}\ \  T_i \le T < T_{i+1},
\label{MHDlamda}
\end{equation}
where

{\begin{tabular}{llll}
$T_1 = 100$     & $L_1= 1.14 \times 10^{15}$ &$b_1=2$\\
$T_2 = 2000$    & $L_2= 5.08 \times 10^{16}$ &$b_2=1.5$\\
$T_3 = 8000$    & $L_3= 2.35 \times 10^{11}$ &$b_3=2.867$\\
$T_4 = 10^5$    & $L_4= 9.03 \times 10^{28}$ &$b_4=-0.65$\\
$T_5 = 4 \times 10^7$.
\end{tabular}}

We refer the reader to Passot et al.\ (1995) and to \VS, Passot \& Pouquet
(1996) for a detailed discussion of this model. Here we just describe it
briefly. The numerical method used for solving these equations, as
well as the purely hydrodynamic (HD) runs discussed in \S \ref{HDmod},
is pseudospectral, implying that diffusion operators have to be
included explicitly, since the method does not produce any numerical
viscosity. The initial conditions are Gaussian with random phases and
a spectrum of the form $P(k)\propto k^4 \exp(-k^2/16)$ in all
variables.  All the simulations 
presented in this section are two-dimensional (2D).

In the momentum equation, eq.\ (\ref{MHDmom}), $\Omega$ is the angular
velocity of Galactic rotation, appearing in the Coriolis term,
and $\nu_8$ is a hyperviscosity coefficient also used in the
magnetic flux freezing equation, eq.\ (\ref{MHDmagn}). 
The usage of hyperviscosity of the form $\nabla^8$ instead of a
standard Laplacian viscosity operator confines viscous effects
to the very smallest scales in the simulations, allowing the development of
larger turbulent inertial ranges. However, it produces oscillations in the
vicinity of strong shocks, as discussed in \S \ref{1Dmod}, and thus a
small amount of second-order viscosity has been added in some runs. 
A mass diffusion term,
with a corresponding coefficient $\mu$, is used in the continuity equation, in
order to smooth out the density gradients, thus allowing the simulations to
reach higher r.m.s.\ Mach numbers. A table with the full list of
fiducial parameter values is presented in Passot et al.\ (1995).

In the internal energy equation, the terms $\Gammad$ and $\Gammas$ respectively
refer to diffuse background and stellar heating rates per unit volume. 
The latter mimics the heating
by ionization occuring in the HII regions surrounding OB stars, which
are modeled as point heating sources appearing at 
every location where the density exceeds a threshold
which we arbitrarily set at a density $\rho_c=30 \langle \rho
\rangle$. The ``stars'' have a lifetime of $6.5 \times 10^6$ yr. Finally,
$\Lambda$ parameterizes the radiative cooling rate per unit
volume. Note that, for compatibility with standard notation, 
we have changed the notation from that in Passot et al.\
(1995) and \VS\ et al.\ (1996). In those papers, $\Gammas$, $\Gammad$
and $\rho \Lambda$ were the heating and cooling rates per unit {\it mass},
and the exponent in the equation defining $\Gammad$ 
was labeled $\alpha$, being related to the exponent used in the
present paper by $a=1-\alpha$.

\begin{figure}[htb]
\plotone{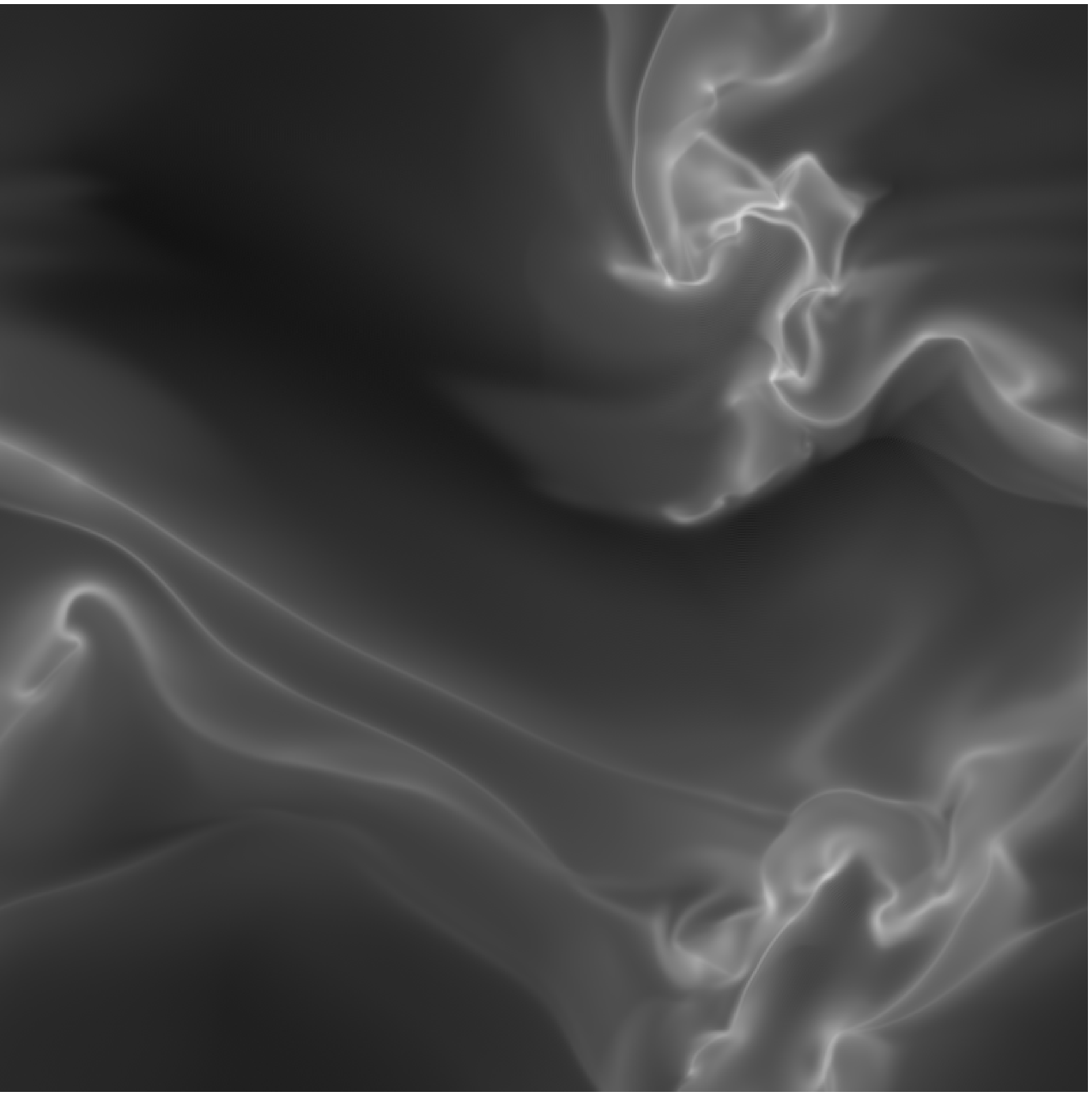}
\caption{Logarithmic grayscale image of the density field of run
  MHD800 at $t=1.2 \to$, where $\to=8.2 \times 10^7$ yr is the sound
  crossing time. The
  simulation represents a region of size 1 kpc, has an initial
  temperature of $10^4$ K, and has an initially turbulent velocity
  field with an rms velocity fluctuation of 11.7 km s$^{-1}$. The 
  maximum and minimum values of the density at the time shown are
  $\rho_{\rm max}=526.3$ and $\rho_{\rm min}=2.42\times 10^{-2}$.}
\label{MHDden}
\end{figure}

Figure \ref{MHDden} shows a typical view of the density field for one MHD
two-dimensional (2D) simulation at a resolution of 
$800 \times 800$ grid points at time $t=1.2 \to$, where $\to=8.2
\times 10^7$ yr is the sound crossing time for the integration
domain. In this run, the dissipation coefficients are 
$\nu_8=5.63\times 10^{-18}$, $\nu_2=\eta=0$, $\mu=3.28
\times 10^{-3}$, and $a=1/2$. The box size 
corresponds to a region of size 1 kpc in
the Galactic plane at the solar galactocentric distance. We refer to
this simulation as MHD800.
It is actually a restart of the run named ``run
28'' in Passot et al.\ (1995), but at larger resolution (run 28 had
$512 \times 512$ grid points). Moreover, the star formation has been
turned off in the last stages of this simulation
in order to allow the turbulence to develop larger density peaks.
Otherwise, the star formation criterion, which turns on a star whenever
the density exceeds $\rho_c$, tends to prevent the simulation from
reaching densities significantly higher than $\rho_c$, since the stars
heat their surroundings, increasing the pressure and producing
expanding bubbles, and reducing the density. This run is discussed in
\VS, Ballesteros-Paredes \& Rodr\'iguez (1997), where another view of
the density field at $t=0.9 \to$ is presented.
Note that the only stellar energy injection
considered is that due to ionization heating of the medium surrounding
OB stars, supernova explosions not being included. The latter implies
that the largest complexes (see below) cannot be blown apart by
the ``stars''.

The density field is seen to consist of dense, filamentary, knotty structures
(``clouds'') which are interconnected in extremely complicated patterns
(``complexes''), being the local peaks
within less dense, extended yet highly amorphous larger structures
(``diffuse clouds'') (e.g., the two large structures in the upper- and
lower-right parts of the figure). In turn, the diffuse clouds gradually
disappear into the lowest-density ``intercloud medium'' (dark regions
in the middle and left upper and lower parts of the image). 

\begin{figure}[htb]
\plottwo{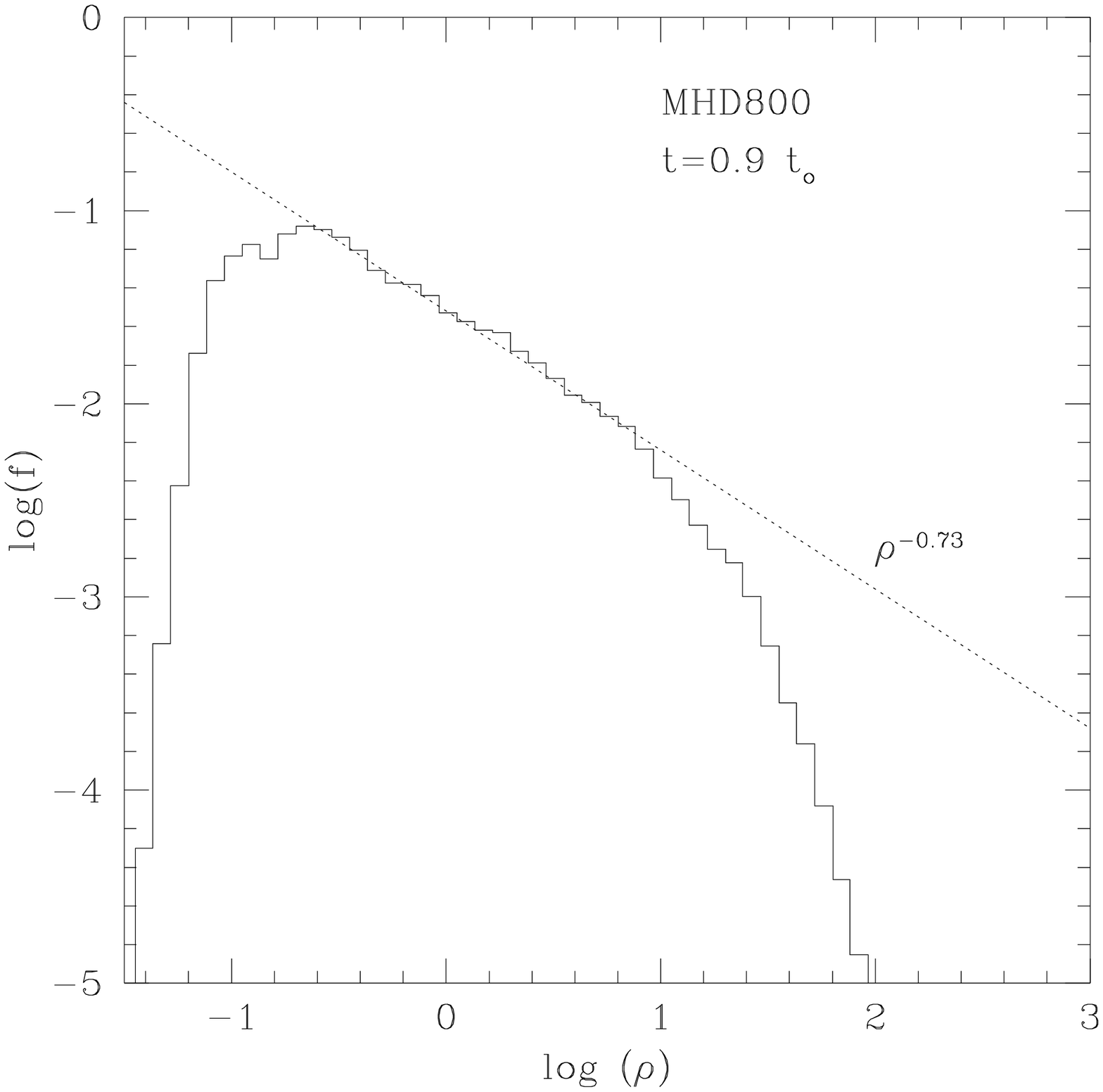}{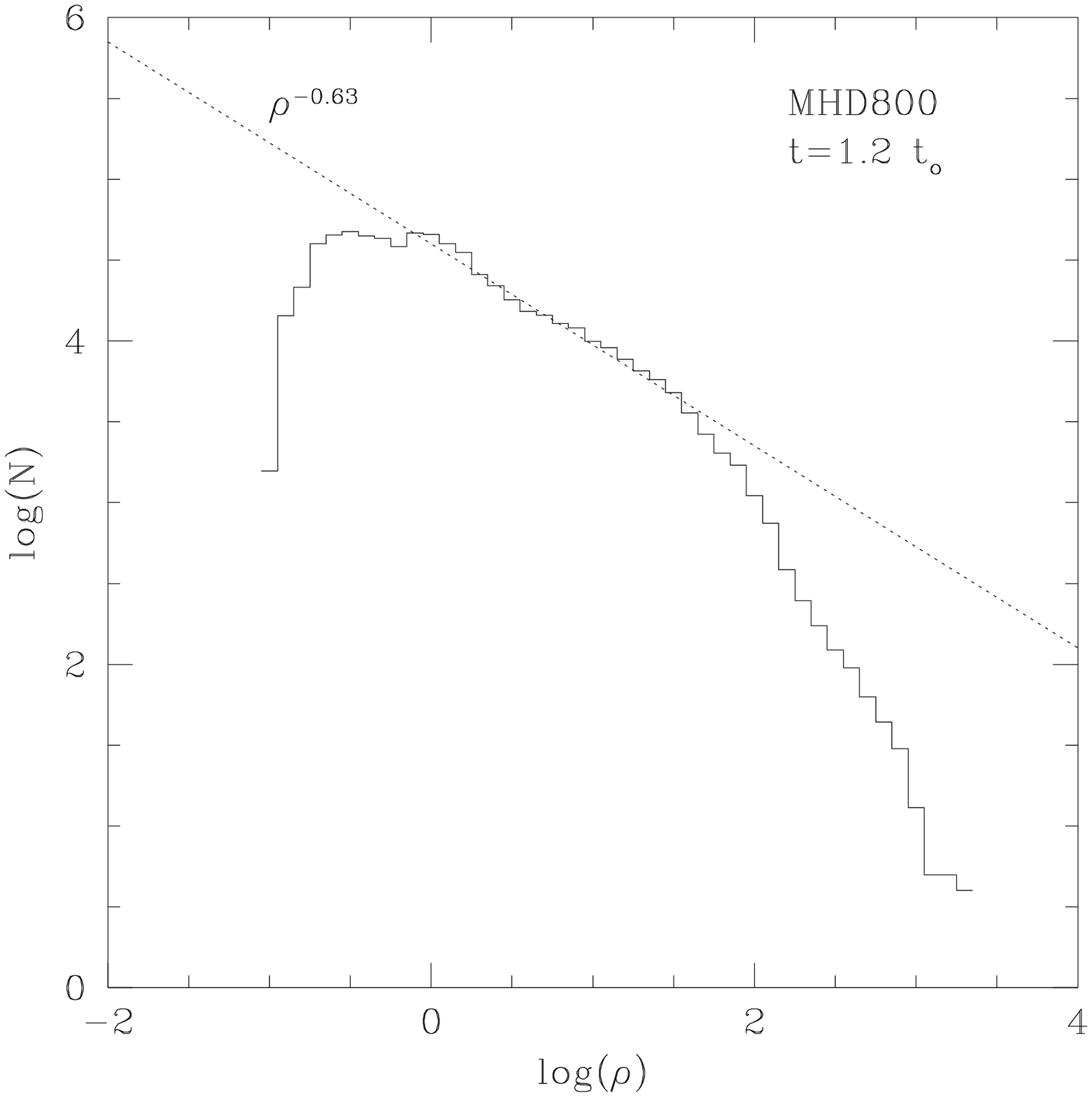}
\caption{Histograms of $\log \rho$  for run MHD800 at
  a) $t=0.9 \to$ (left) and  b) $t=1.2 \to$ (right). The localized
  stellar heating was turned off at $t=0.8
  \to$ in a) and at $t=1.16 \to$ in b). The density field at the latter
  time is shown in fig.\ 1. In a) a clear power law is seen in the range $-0.6
  \leq \log \rho \leq 0.9$, with slope $\sim -0.73$, while in b) a
  slightly less well-defined power law is seen for $0 \leq \log \rho
  \leq 1.5$, with slope $\sim -0.63$.}
\label{MHDpdf}
\end{figure}

Figures \ref{MHDpdf}a and \ref{MHDpdf}b respectively show 
the histograms of $\log \rho$ ($= \rho f(\rho)$, where $f$ is
the density pdf), for run MHD800 at two different times,
$t_1=0.9 \to$ and $t_2=1.2 \to$, the latter corresponding to the
image shown in fig.\
\ref{MHDden}. The density is shown in units of the mean density in the
field. The histogram is computed over the whole field, so it contains
$800^2$ sample points.

The histogram at $t_1$ exhibits a clear power-law behavior in the
interval $0 < \log \rho < 1.3$, with slope $-0.73$. The histogram at
$t_2$, on the other hand, does not exhibit such a clear
power-law behavior, although the region $0 < \log \rho < 1.5$ can be
roughly approximated by a power-law of slope $-0.63$. Both power-laws
are indicated by the solid lines. Note that these slopes correspond to
density pdfs $f$ with slopes 
steeper by one power of $\rho$, i.e., $f \propto \rho^{-1.7}$.

For densities outside the ranges mentioned above, both
histograms turn over, decaying rapidly at both very large and
very small densities. The drop-off at large densities can be easily
understood as a consequence of the viscosity and mass diffusion
included in the equations (the $\nu_8$ and $\mu$ terms in eqs.\
[\ref{MHDmom}] and [\ref{MHDcont}], respectively), whose effect 
is to smooth out tall, narrow
density peaks, ``losing'' them from the histogram. We
speculate that the self-similar range would extend further into
higher density values if more resolution  (and consequently lower
dissipative coefficients) could be used.

The turnover at low densities occurs close to the mean
density, but has a much more abrupt character than either the
Burgers'-type runs discussed in \S\ \ref{Burgers} or the
one-dimensional (1D) simulations
of Passot \& \VS, discussed briefly here in \S\ \ref{1Dmod}. This is
probably due to the presence of self-gravity in run MHD800, which has
managed to collect the gas in large complexes, from which the gas
cannot escape. Therefore, the gas in the voids is only slowly
gravitationally accreted into the complexes, but cannot be swept up by
the filaments as in the Burgers and 1D runs. That is, the density
minima in run MHD800 are probably of gravitational, not turbulent
origin, thus taking much longer times to be evacuated.

\subsection{Hydrodynamic self-gravitating runs} \label{HDmod}

In this section we briefly discuss a first simplification step
down from the full eqs.\ (\ref{MHDcont}) to (\ref{MHDpoisson}),
obtained by eliminating the magnetic field and the Coriolis force due
to Galactic rotation in the simulations. This is an important case,
because these physical agents can be the source of added ``hardness'' in
the flow (\S \ref{gammadisc}; see also \VS\ et al. 1996). 

\begin{figure}[htb]
\plottwo{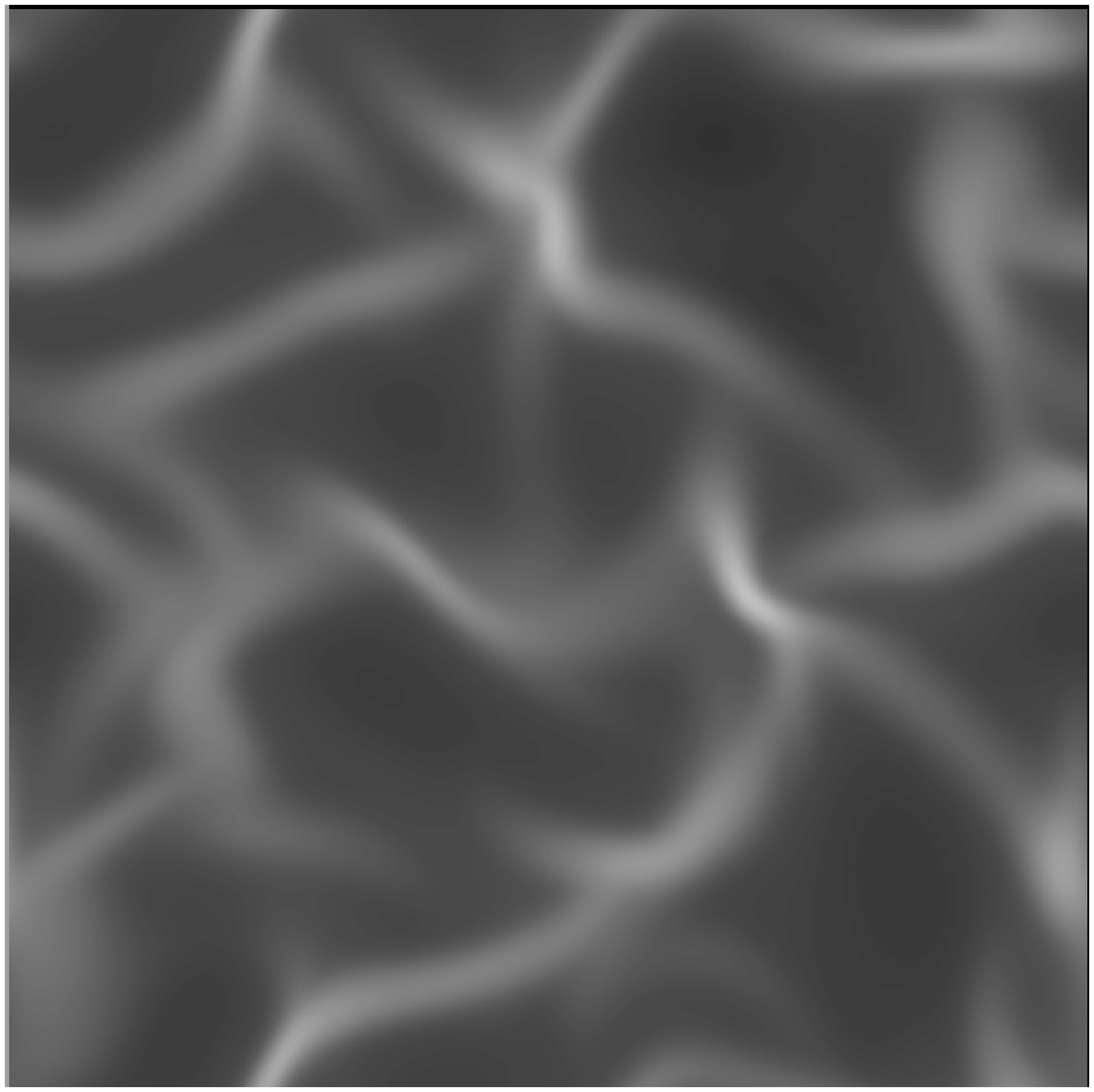}{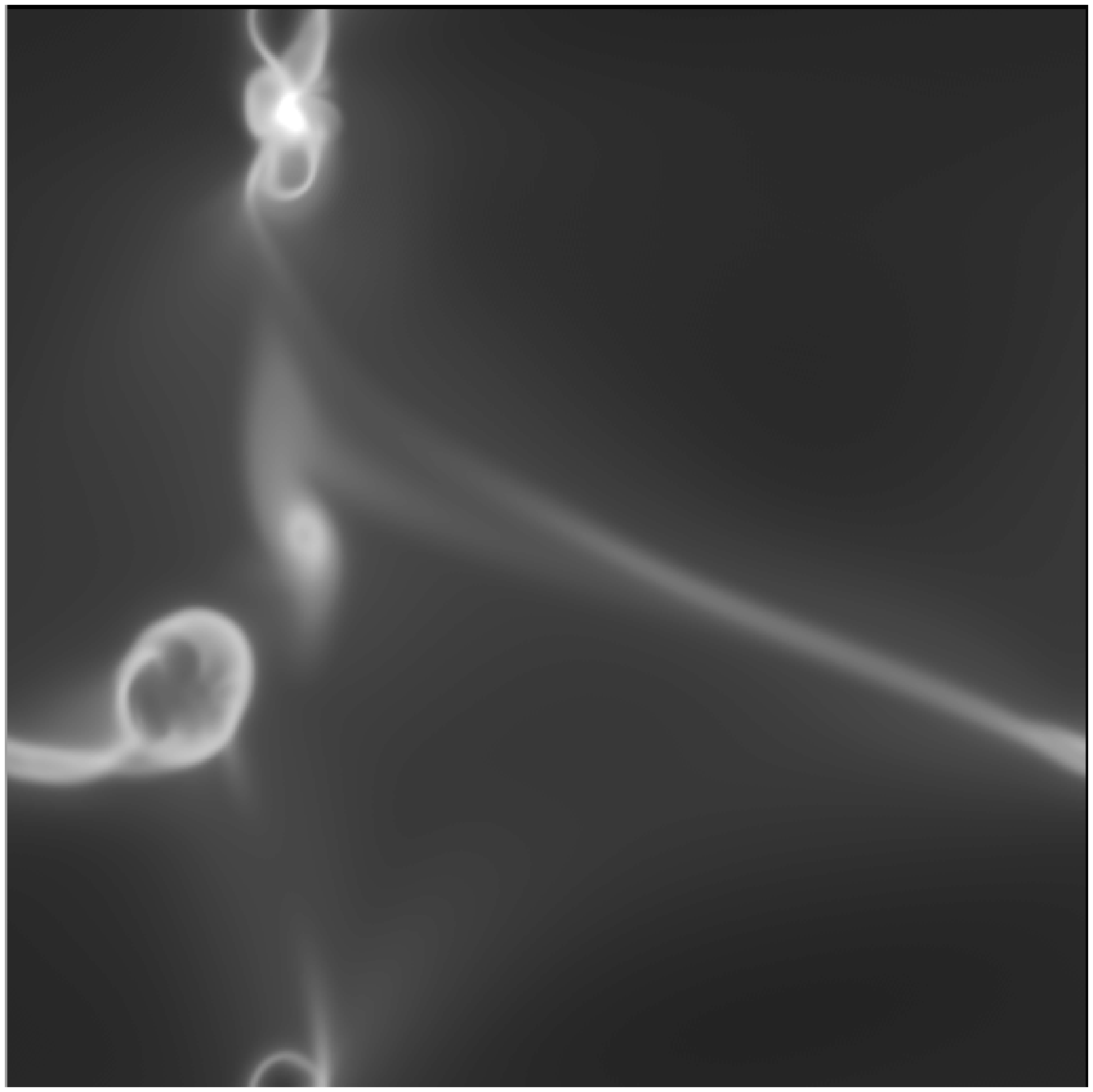}
\caption{Density fields for run HD512 at times a) $t=0.4 \to$ (left) and b)
  $t=1.47 \to$ (right). This is a non-magnetic run otherwise similar to run
  MHD800 (except for the resolution). In a) a network of filaments is
  observed, while in b) material in the upper half of the run has
  collapsed gravitationally. Star formation remained on at
  all times in this run, although it was not able to avoid the
  gravitational collapse observed in b).}
\label{HDdens1}
\end{figure}

Figures \ref{HDdens1}a and \ref{HDdens1}b show 
the density field for a 2D run (labeled HD512) similar to run
MHD800 except without magnetic fields or the Coriolis force, and at a
resolution of 512 grid points per dimension. At this lower resolution,
the values of the diffusion parameters are $\nu_8=10^{-16}$, $\nu_2=2 \times
10^{-3}$, $\mu=8 \times 10^{-3}$, and $a=1/2$. The run considered evolves in
time as follows.  The initial turbulent transients induce the formation
of clumps and filaments, which over the first 0.48 crossing times of
the simulations ($= 3.9 \times 10^7$ yr) merge into
larger and denser structures due to the combined action of turbulence
and self-gravity but with no star formation activity. After this
period, star formation begins, producing isolated expanding ``HII
regions''. However, since no supernovae are included, the stellar
energy input is insufficient to halt collapse of the large structures,
and by $t=1.55 \to$ the largest structure finally collapses. Note that
this type of collapse did not occur in the simulations of \VS, Passot
\& Pouquet (1995), which were also non-rotating and non-magnetic,
because the density threshold for star formation used in that paper
was much lower ($8 \langle \rho \rangle$) and the fluid was harder
($a=1$).

\begin{figure}[htb]
\plottwo{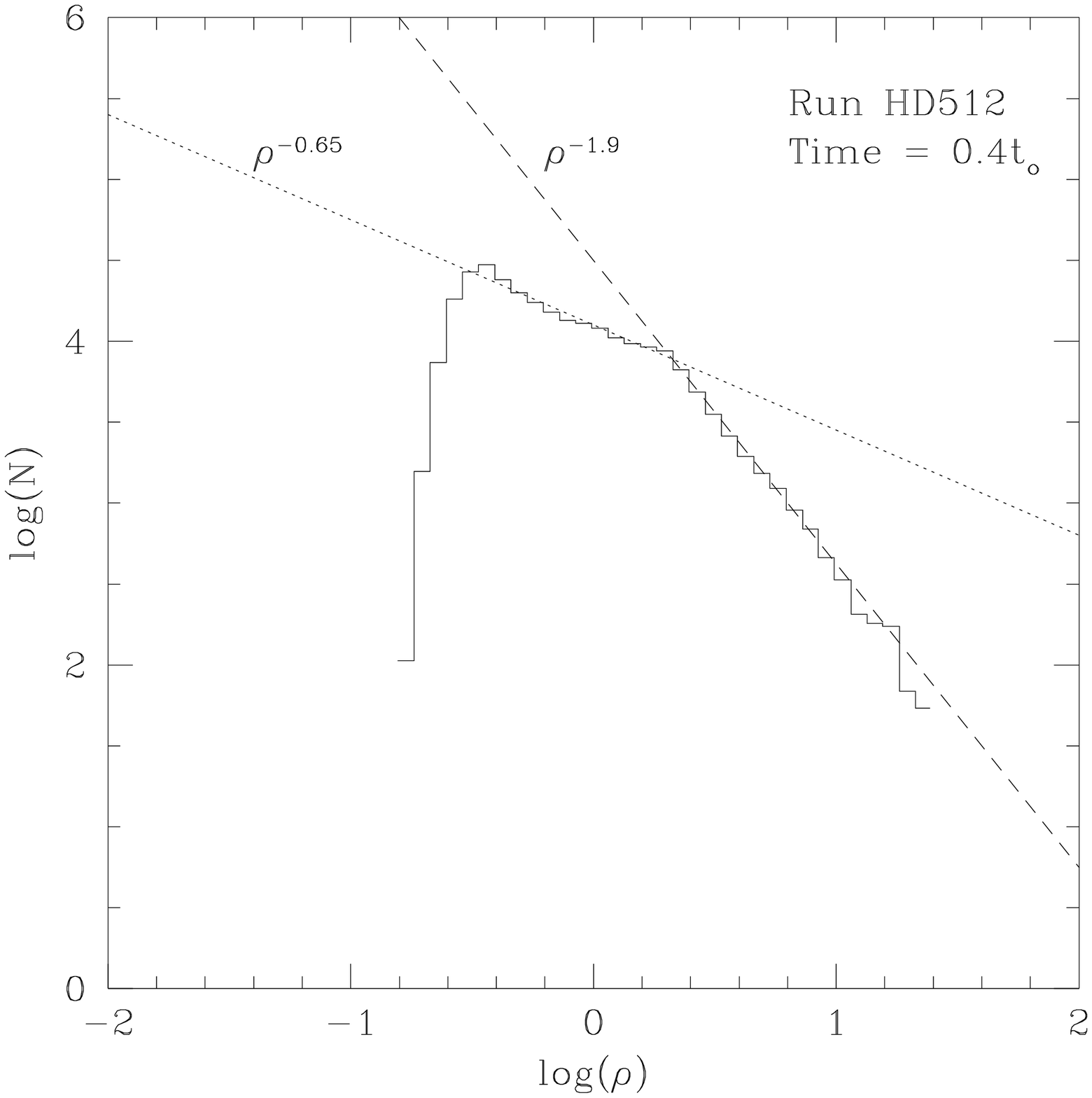}{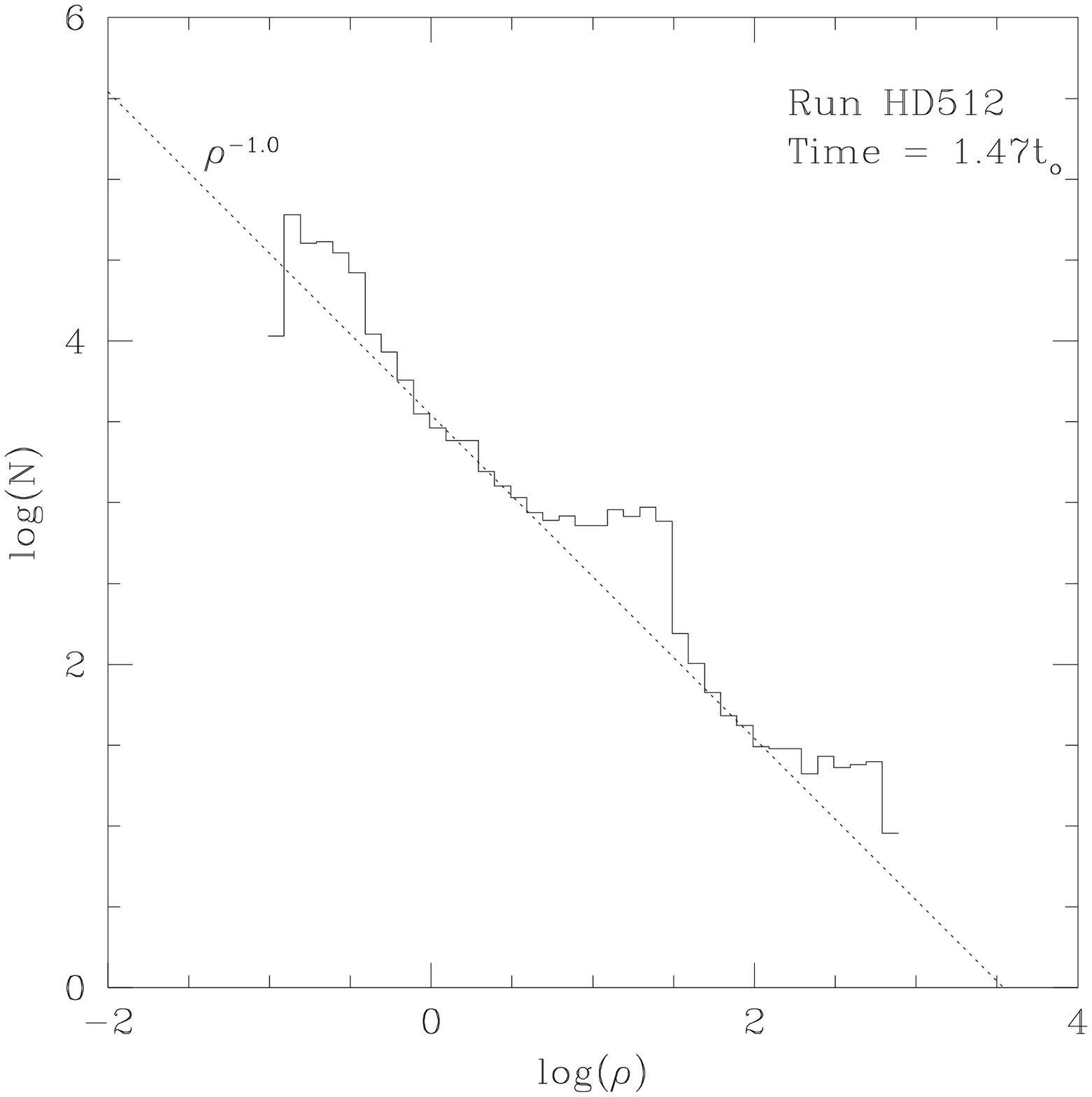}
\caption{Histograms of $\log \rho$ for run HD512 at times a) $t=0.4
  \to$ (left) and b) $t=1.47 \to$ (right), respectively shown in
  figs.\ 3a and b. In a) two power-laws with slopes $-0.65$ and $-1.9$
  can be seen, although the plot can also be interpreted as a single
  power law with a bump at $\log \rho \sim 0.3$. In b) a single power
  law with slope $\sim -1.0$ is observed, with two bumps, one at $\log
  \rho \sim 0.7$ and the other at $\log \rho \sim 2.7$. These seem to
  be respectively due to the onset of star formation at $\rho =30$
  (see text) in the first case, and to the collapsed region, whose
  peak density is $\rho=703$.}
\label{HDpdf}
\end{figure}

The density field is shown at
$t=0.4 \to$ (fig.\ \ref{HDdens1}a) and $t=1.47 \to$ (fig.\
\ref{HDdens1}b). The first time corresponds to an epoch shortly
after the initial transients, at which matter has not significantly gathered
gravitationally, and star formation events have not yet begun. The
corresponding histogram for $\log \rho$ for this epoch is shown in
fig.\ \ref{HDpdf}a, and contains $512^2$ data points. 
Again a near power-law with slope  $\sim -0.6$ 
can be seen  in the range $-0.7 < \log \rho < 0.4$. Interestingly, at
larger densities ($0.4 < \log \rho < 1.5$) another power-law can be
fitted, with slope $\sim -2.0$. Thus, power-law ranges 
in the density pdf appear to be present even after the removal of
magnetic fields and rotation.

The density field shown in fig.\ \ref{HDdens1}b corresponds to the
final state of run HD512, in which a large structure has collapsed
gravitationally in the upper left quadrant of the field. The maximum
density for this field, occuring at the center of the collapse, is
$\rho = 703 \langle \rho \rangle$. The corresponding histogram for
$\log \rho$ is shown in fig.\ \ref{HDdens1}b. Quite surprisingly, even
for this highly singular density configuration, the $\log \rho$
histogram exhibits a power-law range, although with two significant bumps
at $1 < \log \rho < 1.5$ and at $2.5 < \log \rho < 2.75$. These bumps
thus seem to correspond to two special densities: the density
threshold for star formation (recall $\rho_c=30 \langle \rho
\rangle$), and the density of the collapsed region. The bump at
$\rho_c$ can be easily understood since the stellar heating tends to
prevent the density from exceeding $\rho_c$, except when the 
self-gravity of the complex is so large that the stellar heating is
insufficient to prevent generalized collapse. Note again that this
would not occur if supernovae capable of blowing away the complex 
were included. Nevertheless, the
rest of the histogram continues to be a power-law. The measured slope
is $-1$. Thus, the power-law behavior at large densities is preserved
in the absence of magnetic fields and rotation, and even in the
presence of local gravitational collapse events.

\subsection{Stripped-down quasi-Burgers
  simulations} \label{Burgers}

A third series of simulations was run with the intention of stripping-down 
the momentum equation to isolate the effects of the 
advection term $\u\cdot\nabla \u$ in the momentum equation.  For
this purpose the equations 
of continuity and momentum conservation were solved, but all terms on
the RHS of the momentum 
equation were set to zero, except for an additional stellar forcing
term equivalent to that used in the MHD and HD simulations. Both forced
and decaying simulations were performed. Thus
these calculations 
describe a highly compressible fluid in which advection and the
corresponding ``ram pressure" 
completely dominate the thermal pressure, or, equivalently, in which
the effective polytropic index $\gamma$ is zero.  The system is equivalent 
to a forced Burgers flow, except that the 
viscosity in our case is numerical diffusion and the forcing is
coupled spatially to the density field, since star formation and momentum 
input are assumed to occur at a theshold density.  However, the runs
with threshold star formation were 
not capable of generating a sufficiently 
broad range of densities at the high density end to learn much about
the pdf, so we forego any 
discussion of the forced cases and consider only the pure decay
simulations.  The runs with 
star formation (most of the runs) will be discussed elsewhere
(Chappell \& Scalo 1997). 
Because of the similarity to Burgers turbulence, we refer to these
simulations as series ``B," or ``quasi-Burgers" runs.
 
The advection terms were differenced according to a variant of a Van
Leer (1977) first-order 
scheme, altered to eliminate most of the spurious anisotropy of
numerical diffusion in that 
scheme.  The boundary conditions were doubly-periodic.  The initial
conditions were a uniform 
density field and a Gaussian velocity field with prescribed power
spectrum.  Runs with resolution of 128$^2$, 256$^2$, and 512$^2$ were
examined.
The scales were normalized such 
that the lattice spacing was 7.8 pc, so these resolutions correspond to
a total region size of 1, 2, and 4 kpc, although the adopted size
normalization is
probably irrelevant for the results presented here concerning the
density pdf. For pure advection, the hydrodynamic equations are
invariant with respect to a change of scale, so these simulations can
be applied to any scale. 
 
\begin{figure}[htb]
\plotone{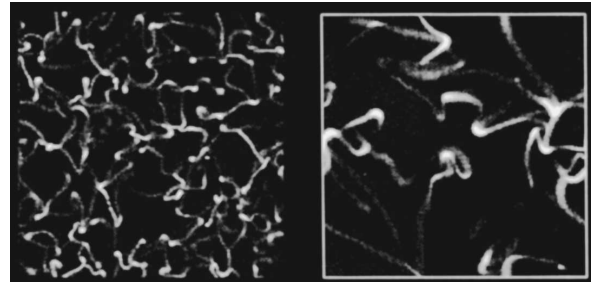}
\caption{
  Spatial distribution of the square root of the density field
  at times 0.1 (left)
  and 10 (right) initial crossing times for a 256$^2$ quasi-Burgers decay
  run.  The simulation was
  initialized with a constant density and a velocity field with energy
  spectrum $E(k)\propto
  k^4 \exp (-k^2/k_o)$ with $k_o=8/L$, where $L=$ size of simulation box.
}
\label{john1}
\end{figure}

All the simulations develop into a network of irregularly shaped
filaments which cover a 
range of sizes.  These filaments are the products of the advection
operator nonlinearly 
self-organizing the initially randomized velocity field into a
collection of shocks.  The 
thicknesses of the filaments are set by numerical diffusion; they are
thinner than in the 
MHD and HD cases because there is no pressure force.  It is just this
absence of pressure 
that ensures the larger compressibilities that feed the
filament-generating advection 
operator.  With no momentum input, the filaments simply continue to
sweep up material and 
each other, so that as time proceeds the structure becomes more
concentrated on large scales 
and the velocities monotonically decrease.  An image of the density
field (actually of 
$\rho^{1/2}$) at two times is shown in Fig.\ \ref{john1}. It is
seen that the structure 
resembles the HD simulation shown in Fig.\ \ref{HDdens1}a at a time
early enough that self-gravity has not 
yet caused the filaments to become concentrated into ``complexes."
 
\begin{figure}[htb]
\plotone{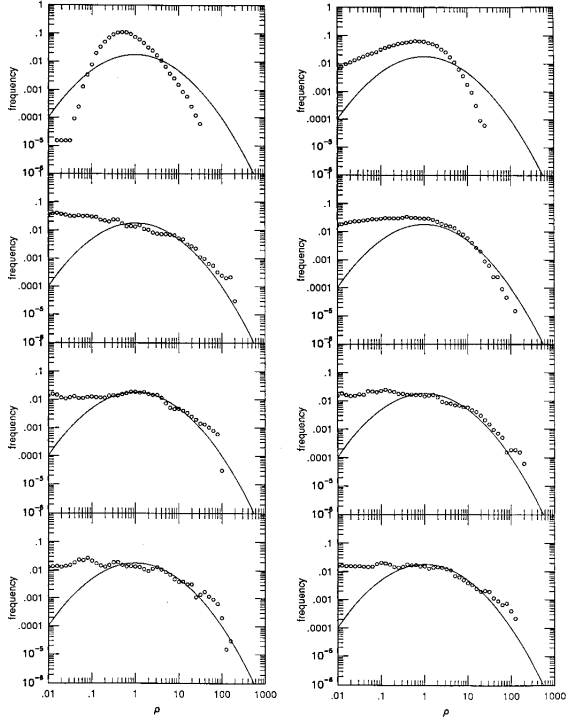}
\caption{
  Time development (top to bottom) of $\rho f(\rho$) for two
  256$^2$ quasi-Burgers
  decay runs.  Initial energy spectra are proportional to $k^{-2}$ (left) and
  $k^0$ (right).  Times,
  from top to bottom, are 0.006, 0.07, 1.1, and 17 initial crossing times.
  The solid lines are
  reference lognormal pdfs with peaks at $\rho=1$.
}
\label{john2}
\end{figure}

Figure \ref{john2} shows the development of the density pdf (actually
$\rho f(p)$) for two sets of 256$^2$ simulations, 
for initial conditions with velocity energy spectra proportional to
$k^{-2}$ (left) and 
$k^0$ (right).  Each pdf is based on about 10$^5$ points, but most of them
are in the voids 
between the filaments.  The solid line is a reference lognormal pdf
with peak at $\rho=1$, 
shown just to illustrate the degree to which the simulation pdfs do or
do not match a 
lognormal.  Times are, from top to bottom, in units of initial
crossing times, 0.006, 0.07, 
1.1, and 17.  After about 1 crossing time one can see the same sort of
power law for log 
$\rho\gta0$ that was found in the MHD and HD simulations.  For both
sets of initial 
conditions the logarithmic slope in the density is about $-0.7$.  The
pdfs become noisy at late times because the density field has evolved
into a relatively small number of large 
scale filaments.  Although the number of points is too small to reveal
the nature of the 
highest-density tail, the dropoff seen in the $k^{-2}$ case, and to a
lesser extent in the 
$k^0$ case, is consistent with the behavior seen in the better-sampled
MHD and HD pdfs; the 
dropoff at highest densities is once again probably due to viscosity,
which smears out the 
smallest scale structures, which are the thinnest, densest filaments
that have recently undergone a collision with another filament.
 
We place no physical significance on the low density ($\rho\lta1$)
part of the pdf.  In this 
pseudo-Burgers flow filaments sweep up all the inter-filament
material, and once the entire 
simulation area has been crossed by at least one filament the density
between filaments 
would be zero except for the action of viscosity (numerical diffusion
here) which causes 
material to leak back into the ``voids."  The low-density pdf extends
to much lower densities than shown in the figure.
 
\begin{figure}[htb]
\plotone{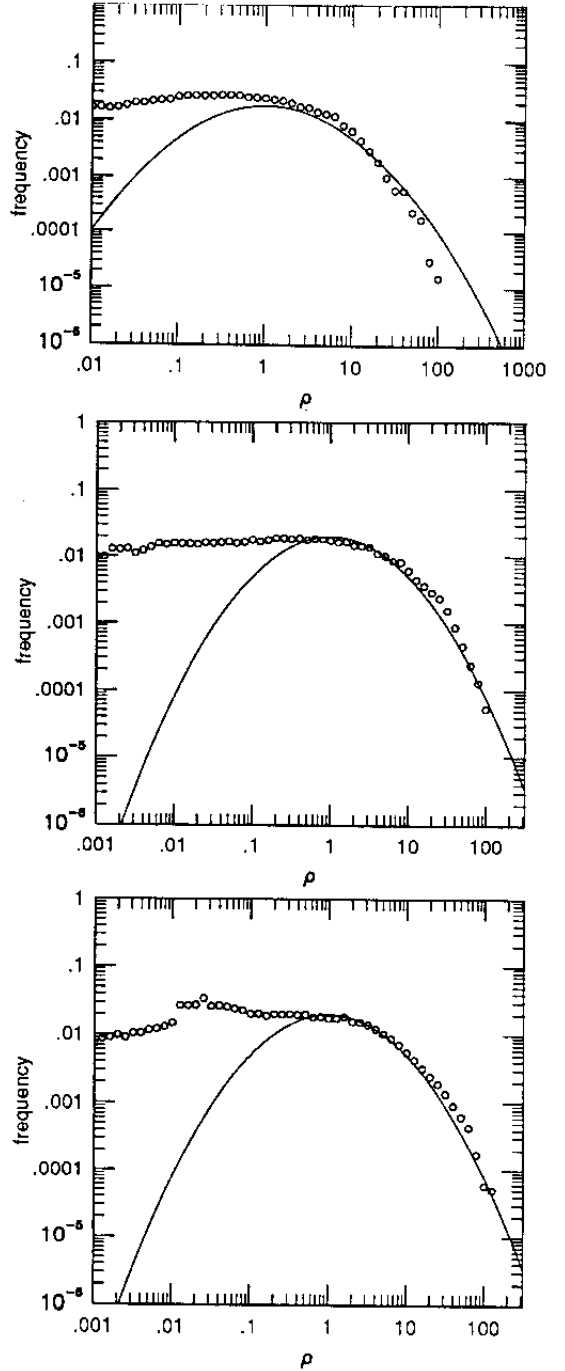}
\caption{
  Late-time distributions $\rho f(\rho)$ for three 256$^2$
  quasi-Burgers decay
  simulations initially excited at a single wavenumber $k_0=64$ (top), 16
  (middle), and 4
  (bottom). Solid lines are reference lognormal pdfs with peaks at $\rho=1$.}
\label{john3}
\end{figure}

Figure \ref{john3} shows the late-time pdfs for three 256$^2$ simulations
initially excited at a 
single given wavenumber $k_0$ = 64, 16, or 4.  Although the impression
is subjective, it 
appears that the pdfs are better represented, for $\rho\gta2$, by a
power law plus a steeper 
dropoff at large densities.  The ``knee" that separates the power-law
density range from the dropoff appears to shift towards lower densities
with increasing k$_0$, occurring at $\rho\approx60$ for $k_0=4$ 
and $\rho\approx30$ for $k_0=16$.  In both cases the logarithmic slope
of the power law is 
around --0.6 to --0.8.  For $k_0=64$ the pdf appears to be fit by the
lognormal for 
$\rho\gta2$, but inspection of other plots for this case at different
times suggests that the 
knee has moved to such small densities, $\rho\sim10$, that the power
law region above 
$\rho\approx2$ spans too small a range to be clearly visible.  The
knee in 
the $k_0=64$ case becomes more prominent at times later than shown in
Fig. \ref{john3}. This is because, 
when the initial velocity fluctuations have such a small scale, it
takes many filament 
interactions to build up large structures.  The same effect can be
seen in Fig.\ \ref{john2} where 
the power law develops much earlier for the $k^{-2}$ initial energy
spectrum than for the 
$k^0$ spectrum.  
 
\begin{figure}[htb]
\plotone{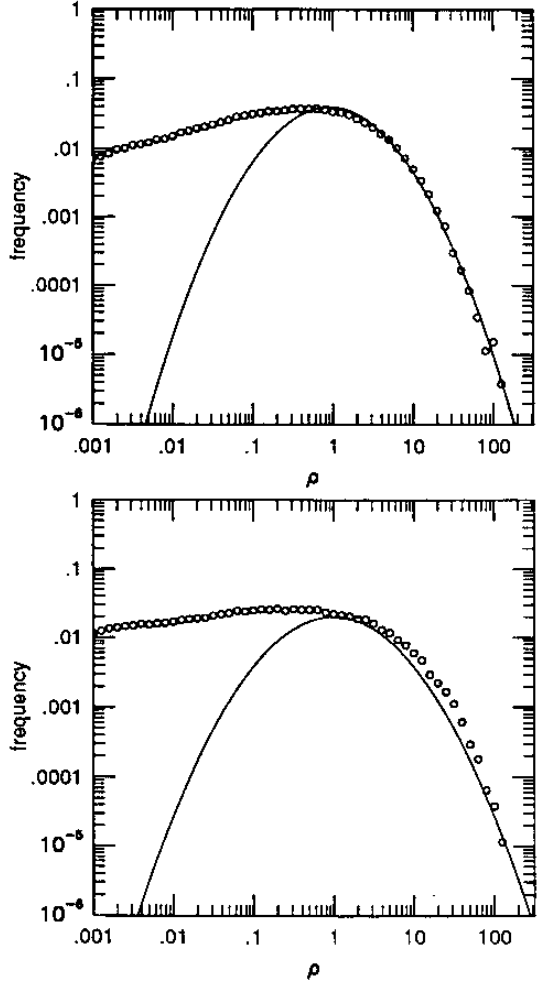}
\caption{
  Distribution $\rho f(\rho)$ for a 512$^2$ quasi-Burgers decay
  simulation at times
  0.1 (top) and 2 (bottom) initial crossing times.  Solid lines are reference
  lognormal pdfs with
  peaks at $\rho=1$.
}
\label{john4}
\end{figure}

In order to increase the sampling of cells at large densities, we ran
a 512$^2$ simulation 
with a broad-band initial spectrum.  The resulting density pdf is
shown in Fig.\ \ref{john4} at two 
times.  At early times the pdf at densities greater than the average
appears to fit a 
lognormal, but after about 1.5 crossing times the result is more
suggestive of a short power 
law segment with a falloff at $\rho\gta10$.  Between log$\rho=0.2$ and
1.0 the logarithmic slope is about --0.6.

We note that Gotoh and
Kraichnan's (1993) 1-dimensional simulations of the decaying Burgers equation
produced a density pdf with a clear power-law form, although steeper than
found here. Their power law regime had a larger extent in
density probably because of 
the much greater dynamical range possible in 1D compared to the present 2D
simulations. We take their work as support for the power law pdf regimes
suggested here.
 
Because of the poor sampling statistics for these quasi-Burgers decay
simulations compared to the MHD and 
HD runs, the results are not conclusive, but the pdfs are generally
consistent with the power 
law segment plus high density falloff found in the more physically
complete MHD and HD 
simulations.  The rough agreement is especially surprising because the
B simulations used a 
very different numerical method and, in most cases, different initial
conditions.  Since the 
B simulations have omitted all the physics except for nonlinear
advection, this rather 
surprising agreement suggests that {\it the physical process
  responsible for the power law portion of the 
  pdf in the high-density side is nonlinear advection.}  Intuitively,
this can be understood in 
terms of the tendency of the advection term to induce multiplicative
processes which may lead to power laws, if the equation of state
permits it (Passot \& \VS\ 1997, hereafter PVS).
 
 \section{The Role of the Polytropic Index}

\subsection{One-dimensional simulations}\label{1Dmod}

The simulations discussed in the previous sections suggest (with varying
degrees of precision) the development of power-law regions at high
densities in the density pdf of interstellar
gas. This result is in contrast with previous work (\VS\ 1994;
Padoan et al.\ 1997a,b,c) finding lognormal distributions in a
variety of numerical simulations. \VS\ (1994) presented isothermal
($\gamma=1$) simulations of two-dimensional, randomly-forced 
Navier-Stokes turbulence in the weakly
compressible regime. No other processes, such as star formation,
self-gravity, magnetic fields, etc., were included. Padoan et al. 
(1997a, b, c) refer to highly-compressible three-dimensional
isothermal MHD simulations with random forcing and without  
self-gravity to be presented elsewhere,  although, to our knowledge,
the pdfs have not been published yet. For both sets of
simulations, a lognormal pdf is reported.

The discrepancy between our pdfs and those of \VS\ (1994) and
Padoan et al.\ (1997a,b,c) can be understood in terms of some recent
results by PVS, which we briefly
summarize here. As a first step towards understanding the relation
between the dynamical and statistical properties of highly
compressible turbulent flows, PVS have performed a systematic
investigation of one-dimensional (1D) simulations of pure
Navier-Stokes polytropic turbulence (with mass diffusion added on some
runs in order to allow the simulations to tolerate strong shocks), 
finding that {\it a lognormal density pdf
only appears when $\gamma=1$}. In fact, as explained below,
PVS suggest that the case $\gamma=1$ may be singular in this respect. 

The equations solved by PVS in nondimensional form are:
\begin{equation}
{\partial\rho\over\partial t} + {\partial(\rho u) \over \partial x} = \mu
{\partial^2 \rho \over \partial x^2}, \label{1Dcont}
\end{equation}
and
\begin{equation}
{\partial u\over\partial t} + u{\partial u \over \partial x} =
-{1 \over \gamma M^2\rho }{\partial \rho^{\gamma} \over \partial
x} + \nu_2 \bigl({\partial^2 u \over \partial x^2}\bigr) + a_r,\label{1Dmom}
\end{equation}
where $M$ is the Mach number of the velocity unit. In these
simulations, only  second-order viscosity is used in order to avoid
the spurious oscillations induced in the vicinity of shocks by the use of
hyperviscosity (Passot \& Pouquet 1988). Mass diffusion (the RHS of
eq.\ [\ref{1Dcont}]) is used in small amounts to help the
simulations survive strong shocks. The probability distributions of
$\log \rho$ are obtained by considering all grid points (typically
2048) and averaging over nearly 150 code non-dimensional time units,
sampling every 0.1 time units. Thus, the histograms contain over 3
million points.

\begin{figure}[htb]
\plotone{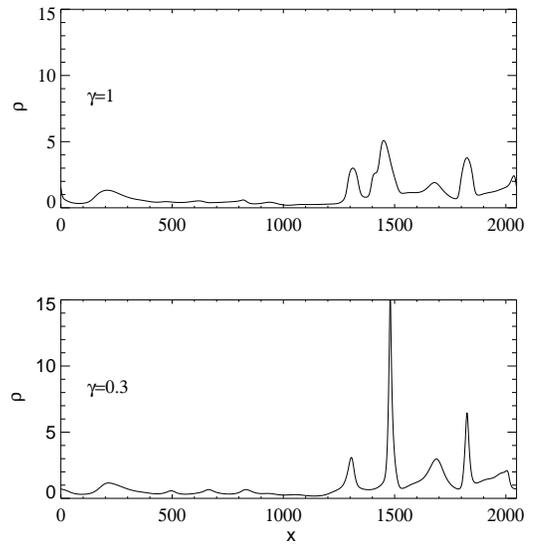}
\caption{Density fields for two one-dimensional runs with $M=3$ (see
  eq.\ [12]), with a resolution of 2048 grid points. a) $\gamma=1$
  (top) and b) $\gamma=0.3$ (bottom). These 
  runs all start from rest, but are forced with white noise at the
  large scales (see text) and with a correlation time $t_{\rm corr}
  =4.77 \times 10^{-4}\to$. The run with $\gamma=0.3$ shows taller,
  narrower density peaks than that with $\gamma = 1$.}
\label{1Dden}
\end{figure}

\begin{figure}[htb]
\plotone{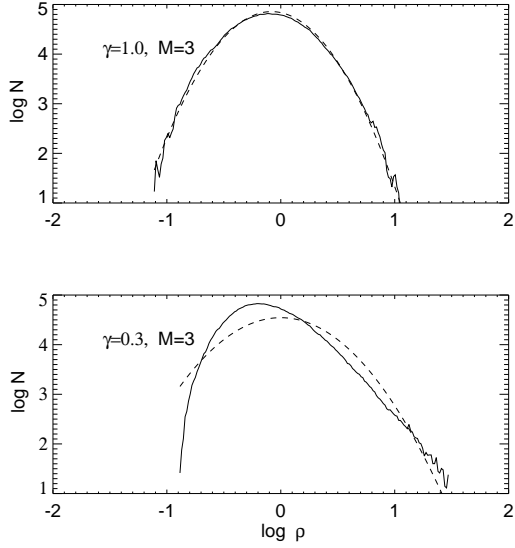}
\caption{Histograms of $\log \rho$ for the 1D simulations shown in
  fig.\ \ref{1Dden}, with $M=3$. The runs are sampled at intervals
  $\Delta t=1.59 \times 10^{-2} \to$ over a total time $t=23.9
  \to$. The histograms thus include $1500 \times 2048$ data
  points. The dotted lines show
  lognormal fits to the curves. a) $\gamma=1$ (top). The fit by a
  lognormal is excellent. b) $\gamma=0.3$ (bottom). The histogram
  differs drastically from a lognormal, and instead a clear power-law
  is seen at high densities, in the range $0.3 \leq \log \rho \leq
  1.3$.}
\label{1Dpdf}
\end{figure}

\begin{figure}[htb]
\plotone{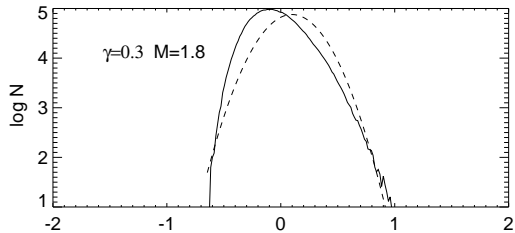}
\caption{Histogram of $\log \rho$ for a 1D simulation similar to that
  shown in fig.\ \ref{1Dden}a, except with $M=1.68$. The deviation
  from a lognormal is seen to be less pronounced than in the case
  $M=3$. Also, the power-law at high densities is less
  well-developed.}
\label{1Dpdf_Ma}
\end{figure}

As an illustration, we show in figs.\ \ref{1Dden}a, and b the
density fields of two different simulations, both with $M=3$, and
$\nu_2=\mu=0.003$, but with $\gamma=1$ in (a) and $\gamma=0.3$ in
(b). These runs start at rest, but are driven with a random
acceleration $a_r$ at
wavenumbers 2--20, with a correlation time $t_{\rm corr}=4.8 \times
10^{-4} \to$.
The associated histograms for $\log \rho$ are
shown in figs.\ \ref{1Dpdf}a and b. The dotted lines show a
least-squares fit to a lognormal 
curve. As can be seen, for the simulation with $\gamma=1$
the fit is excellent, suggesting that indeed the $\log
\rho$ histogram is a true lognormal. Instead, in
figure \ref{1Dpdf}b, a clear power-law tail is seen to develop
at large density fluctuations.
Furthermore, in fig.\ \ref{1Dpdf_Ma} we show the corresponding
histogram for a run with $\gamma=0.3$
but $M=1.8$. Upon comparison with fig. \ref{1Dden}b, this run
illustrates the effect of 
varying the Mach number $M$ at a fixed $\gamma$. The deviation from a
lognormal towards a high-density power-law tail is more noticeable at
large $M$. These results are consistent with the power-law functional
form of the pdf at high densities found
by Gotoh \& Kraichnan (1993) in their 1D numerical simulations of decaying
Burgers turbulence, which effectively have $\gamma=0$. Also, note that
the development of the high-density power-law tail in the pdfs of the present
simulations cannot be attributed to a numerical effect, such as the
inclusion of the mass diffusion term in the continuity equation. Runs at lower
Mach numbers (not shown), which could be performed without the need
for such a term, still exhibited a power-law tail, although not as
well developed, due to the Mach number dependence discussed above.

In order to interpret these results,
it is instructive to rewrite the inviscid one-di\-men\-sion\-al gas dynamics
equations in Riemann invariant form. The
Riemmann invariants of the problem are defined, for $\gamma \ne 1$,
 as $z^{\pm}=u\pm Kc$,
where $u$ is the one-dimensional fluid velocity, $c= \rho^{(\gamma-1)/2}/M$
is the sound speed, and $K=\pm 2/(\gamma-1)$. 
They are just advected
along the characteristics whose speeds are $u\pm c$. 
In the case $\gamma=1$, $z^{\pm}=u\pm \log\rho/M$ while the 
characteristic speeds are $u\pm 1/M$. 
One can easily see that, under the transformation $\rho\rightarrow
1/\rho$, $\gamma\rightarrow 2-\gamma$, the characteristic speeds
remain unchanged while the Riemann invariants are exchanged.
Under such a transformation the local dynamics is not preserved but
the statistical quantities such as the pdf
remain close. For $\gamma=1$ this indication is consistent with
the fact that the pdf is a lognormal, a curve symmetric under the
change $\log\rho\rightarrow -\log\rho$.

It should be noted that the lognormal pdf at $\gamma=1$ disappears
when a random force $f_r$ is
used, leading to replace $a_r$ by $f_r/\rho$ in eq.\
(\ref{1Dmom}). We speculate that this is due to 
the strong nonlinearity introduced by the division by $\rho$. Thus, to
study the effect of the advection operator on the density pdf, the
usage of an acceleration seems more adequate.

In terms of these results, we can interpret the previous
claims of lognormal pdfs (\VS\ 1994; Padoan et al.\ 1997a,b,c) in
a simple way. Since both sets of simulations were isothermal, their
resulting pdfs were bound to be lognormal, independently of the Mach
number. However, for any other values of the polytropic exponent,
power-law pdfs should be realized.

\subsection{The polytropic index of cool interstellar gas} \label{gammadisc}
 
The one-dimensional calculations of supersonic \break\hfill Navier-Stokes 
turbulence discussed above and presented in detail by 
PVS strongly suggest that the form of the 
density pdf $f(\rho$) depends sensitively on the effective
polytropic index $\gamma$.  In 
particular, if the turbulence is highly supersonic, $f(\rho)$ is
strictly lognormal only for 
$\gamma$ equal to unity, while for smaller (larger) $\gamma$ a power
law develops for 
densities larger (smaller) than the mean.  The sensitivity of hydrodynamic
behavior to $\gamma$ 
has been found previously in non-turbulent contexts.  For example,
Pongracic (1994)
and Foster \& Boss (1996) showed that the ability of a shock incident on a
model cloud to induce 
gravitational collapse depends on $\gamma$ (through a piecewise power
law cooling rate for 
Pangracic).  \VS\ et al.\ (1996) estimated a critical
value of the polytropic exponent below which
turbulence can induce bound condensations, as a function of the
dimensionality of the turbulent compressions.
 
The polytropic index $\gamma$ defined by $P\sim\rho^\gamma$, is identically
zero for the 
quasi-Burgers simulations.  For hydrodynamic simulations which allow
for heating and 
cooling, $\gamma$ can be determined locally by the temperature and
density dependence of the 
heating and cooling function.  If the heating and cooling rates per
unit volume depend on 
temperature and density according to $\Gamma\sim\rho^a$ and
$\Lambda\sim\rho^cT^b$, then the 
condition of thermal equilibrium (which holds for the flows of
interest because of the small 
cooling times, except for just behind shocks), gives
\begin{equation} 
\gamma=1+\frac{\partial \log T}{\partial
\log\rho}=1-\frac{(c-a)}{b}\left(1-\frac{\log\rho}{b}\frac{\partial b}
{\partial \log\rho}\right) \label{gamma}
\end{equation}
(cf. Elmegreen 1991; Passot et al. 1995; \VS\ et al.
1996).  The last factor in parentheses allows for the possibility that
the temperature 
dependence exponent $b$ in the cooling function is itself a continuous
function of the density (see eq.\ [\ref{temp_exp}] below).
 
In the MHD and HD simulations discussed above, the heating was assumed
to be due to a 
combination of stellar  and diffuse radiative heating, with
$a=0.5$ for the latter, while the cooling 
rate was a piecewise power-law function appropriate to temperatures
larger than a few 
hundred degrees, because the simulations were designed to model
large-scale dynamics of 
the atomic gas.  For those simulations the values of $\gamma$ are in the range
0--0.5.\footnote{Actually, the range of cooling functions
  described in sec. \ref{models} 
 give $\gamma=3.3$ for $10^5\leq T<4\times10^7$ K.  However, those
 temperatures are never 
 reached in the simulations.}
Therefore the development of the power law $f(\rho$) at large densities
found in the MHD, HD, 
and B simulations is consistent with the one-dimensional results,
since $\gamma$ is 
significantly smaller than unity.
 
However, the most interesting applications of the density pdf concern
regions that can form 
stars, which are at small temperatures and are generally molecular instead
of atomic.  We 
therefore need to estimate the value of $\gamma$ in cool molecular
gas.  A few previous 
attempts in this direction are summarized by Larson (1985, Fig. 2) who suggests
$\gamma\approx0.7$ in the absence of stellar heating, for $T\sim10-50$ K and
$n\sim10^1-10^5$ cm$^{-3}$.  However there are a number of possible
heating mechanisms, and 
the cooling is complicated by the contribution of different coolants
and optical depth effects at large densities.
 
A major study of interstellar cooling and heating at low
temperatures ($<100$K) was given by Goldsmith and Langer (1978, hereafter GL)
for the density range $10^2$--$10^5$cm$^{-3}$. The resulting cooling function
is dependent on the prescribed abundances of coolants and the treatment of
radiative transfer (escape probability method with constant large velocity
gradient).  A considerably more detailed treatment of the cooling function
was presented by Neufeld, Lepp, and Melnick (1995, hereafter NLM) for
densities $10^3$--$10^{10}$cm$^{-3}$ and temperatures 10--2500K.  Besides
including more coolants, using updated data on cross sections, and a
modified velocity gradient parameter for the radiative transfer, NLM
obtained coolant abundances by a self-consistent (but steady-state)
solution to a large chemical reaction network. The results assume complete
shielding from ultraviolet radiation, and may be
significantly different in lower-column density regions.
        We base the following discussion of $\gamma$ on the results of GL and
NLM, concentrating on densities $10^2$--$10^5$ cm$^{-3}$ and
temperatures 10--100 K. 
We emphasize that our purpose is not to give a comprehensive treatment of
the cooling and heating functions, but only to outline what seem to be the
dominant physical effects on $\gamma$ and to motivate a more detailed
treatment.

For temperatures less than about 100 K, we find that
the temperature dependence of the cooling 
function, parameterized by $L \sim T^b$, can be adequately 
represented by a density-dependent function, based on Table 4 of GL,
\begin{equation}
b=\frac{1}{2} (1+ \log\ n) \label{temp_exp}
\end{equation}
for $\log\ n$ in the range 2 to 5, where $n$ is the total particle
density.  The numerical and graphical
results presented by NLM are not sufficient to confirm this relation,
although the trend of larger temperature dependence at larger densities
seems to hold for densities up to $10^5$ cm$^{-3}$, based on Fig.2 in NLM. At
smaller densities, based on the data in GL, $b$ is smaller than given by eq.\
(\ref{temp_exp}). For example, at $n=10^1$ cm$^{-3}$, $b\approx 0.5$.
 
For densities less than a few hundred cm$^{-3}$ the total cooling rate
per unit volume 
varies with the square of the density ($c=2$), but at densities around
10$^3$ cm$^{-3}$ or 
larger, the density dependence becomes much weaker because of
radiative trapping (e.g.\ GL).  The density at which the
flattened density dependence occurs depends on a number of effects, but is
about $10^3$cm$^{-3}$ for the parameters adopted by GL.
The resulting density dependence is
uncertain because 
of the approximate nature of the radiative transfer calculations in
the available work and 
the assumption of a smooth (not turbulent) velocity field.  Inspection
of Fig. 3 of NLM indicates that for $n$ between 10$^3$ and 10$^5$
cm $^{-3}$, $c$ is 
about 1.6 for $T=100$ K and 1.2 for $T=10$ K, although at $T=10$ K $c$
is very close to 
1.0 for their singular isothermal sphere model.  For reasons we do
not completely understand, the 
density dependence exponents at high densities are smaller by about 0.6 in the
calculations of 
GL, Figs. 7a--c, with $c$ as small as 0.5 at
$T=10$ K.  Some possible reasons for the differences between GL and
NLM are given by NLM, although they do not explicitly discuss the
differences in density dependence.
For illustrative purposes we will adopt $c=1.2$ for $n>10^3$ cm$^{-3}$
and $T=10$ K. 
 
The heating rate is problematic because it is uncertain what the
dominant process is (see GL, Black 1987, and Hollenbach 1988 for
reviews).  It is often 
assumed that the main heating agent is cosmic rays, with
$\Gamma\sim\rho$, so that $a=1$. 
The heating due to H$_2$ formation on grains depends on the fraction
of neutral hydrogen, 
but at large densities the heating rate is again proportional to
$\rho$.  Heating rates 
due to gas-grain collisions, compression or collapse, ambipolar 
diffusion, or turbulence are more complicated and uncertain, and may include a
temperature dependence.  We ignore these sources here, except to note
that $b$ may be 
significantly larger than unity for these cases. An extreme case is
gas-grain heating, which is 
proportional to $\rho^2T^{1/2}(T_{gr}-T)$, where $T_{gr}$ is the grain
temperature. This case is discussed below.
 
Our results suggest different behavior of $\gamma$ in four density ranges.

1.  $n<10^3$ cm$^{-3}$.  In this case, taking the density dependence
of the heating as $a=1$, and including the  density dependence of
$b$ (eq.\ [\ref{temp_exp}]) in eq.\ (\ref{gamma}), an effect that
pushes $\gamma$ closer to unity,  
the polytropic index is 0.56 for 
$n=10^2\ (b=1.5)$ and 0.88 for $n=10^3\ (b=2)$.  These bracket the 
estimate by Larson (1985).  At densities as low as 10 cm$^{-3},\
b\approx0.5$ based on the GL cooling rates, and $\gamma$ 
is negative, corresponding to thermal instability.  
 
We conclude that, as long as the dominant heating mechanism is cosmic
rays, H$_2$ 
formation, or some other process whose rate scales with the density in
a manner not much 
steeper than linearly, the polytropic index will be sufficiently
smaller than unity in 
molecular regions with $n<10^3$ cm$^{-3}$ that the power law density
pdfs found in the 
simulations should apply there, also.  A lognormal pdf should only
obtain for a nearly 
quadratic density dependence of the heating rate per unit volume.

2.  $n\sim10^3$ to a few times $10^4$ cm$^{-3}$.  The situation is 
different at densities larger than 10$^3$ cm$^{-3}$, at least when 
gas-grain collisional cooling does not dominate, 
because radiative trapping weakens the density dependence of the cooling rate,
i.e. reduces the exponent 
$c$.  At $T=10$ K, adopting $c=1.2$ from the results of Neufeld et
al. (1995) and again 
using $a=1$ for the density dependence of the heating rate, $\gamma$
varies from 0.98 at 
$n=10^3$ to 0.99 at $n=10^5$.  At $T=100$ K, taking $c=1.6,\ \gamma$
varies from 0.92 to 
0.97 over this density range.  The polytropic indices are close enough
to unity that the lognormal 
density pdf might occur. There is some inconsistency
in arriving at these numbers, because the density dependence of the
temperature exponent in the cooling rate (eq.\ 14) was based on GL, while the
density exponent is based on NLM.  Without the density dependence of $\gamma$,
the derived values of $\gamma$ are slightly smaller, but still close to unity.

That $\gamma$ may be very close to unity at densities above $10^3$
cm$^{-3}$ is by no means a 
definitive result.  After the above calculations were done, we found
that Lis \& Goldsmith 
(1990) have given polynomial fits to the cooling function of GL, which 
includes radiative trapping.  Taking analytical derivatives of their
expressions in order to 
evaluate $\gamma$, and assuming a linear dependence of the heating
rate on density, we find that 
at log $n=(2,3,4)\ \gamma=$ (0.89, 1.21, 1.26).  This suggests that
$\gamma$ will only be very 
close to unity in a narrow density range around 300 cm$^{-3}$.  The
reason for the difference is 
that the effective values of $c$, the density dependence of the
cooling rate, are continuously 
varying and significantly smaller than the value $c=1.2$ which we
estimated for the high-density 
cooling functions of NLM.
 
Based on the NLM cooling rate, we suspect that a major transition occurs
at densities above 
which radiative trapping in the cooling lines becomes important.  At
smaller densities the 
polytropic index should be significantly smaller than unity, and a
power law density pdf 
should occur.  At larger densities the value of $\gamma$ should be
close to unity, and a 
lognormal density pdf is possible.  Since these different density pdfs
are a reflection of 
the velocity field (which generates density fluctations through the
dilatation $\nabla\cdot 
\u$), we expect that the velocity field and other associated phenomena
should also be qualitatively different in the two density regimes.  
One caveat to this conclusion is that radiative trapping may be very 
different for a turbulent velocity field than for the linear velocity 
gradients assumed in the existing cooling rate calculations. (For a
calculation of 
molecular line formation in a turbulent velocity field, but without density
fluctuations, see Kegel, Piehler \& Albrecht
1993.)  Another caveat is the fact that we obtain values of $\gamma$
significantly larger than 
unity for $n = 10^3$--$10^4$cm$^{-3}$ using the fits of Lis \& Goldsmith
(1990) to the cooling function 
of GL, as discussed above.

The possibility that $\gamma$ may be close to unity at densities
between 10$^3$ and 10$^4$ 
cm$^{-3}$ is consistent with the near constant temperature $\sim 10$ K
observed in  
dark clouds without internal protostellar heat sources. However, this
latter result is based 
largely on CO observations that sample a fairly narrow range of
densities.  Often this result is 
ascribed either to the efficiency or the temperature dependence of the
cooling; however it is 
clear that near isothermality should occur in this density range
primarily because the density 
dependence of the cooling rate changes as radiative trapping becomes
important.

3.  $n\gta10^4$ cm$^{-3}$.  At still larger densities, a different
physical effect becomes 
important.  If there are no embedded protostellar sources to heat the
dust grains, then cooling 
by gas-grain collisions should dominate the molecular cooling at
densities above $1\times10^4$ 
cm$^{-3}$ to
$5\times10^4$ cm$^{-3}$, depending on the grain temperature and grain
parameters.  (For the gas-grain energy transfer rate, see, for
example, GL and Hollenbach and McKee 1989.) The gas-grain 
cooling rate is proportional to $\rho^2T^{1/2}(T-T_{gr})$, so $c=2$
and $b\approx3/2$ if $T\gg 
T_{gr}$.  In this case $\gamma\lta1/3$ and so a return to a power law
density pdf should occur 
for densities above
$(1-5)\times10^4$ cm$^{-3}$.  However the grain, and hence gas,
temperature is limited by the 
cosmic background radiation to $T\approx3K$, so it may be difficult to
detect the decrease in 
gas temperature in this density range.

If embedded protostellar sources are present, the grain temperature
will probably exceed the gas 
temperature and gas-grain collisional heating will dominate cosmic ray
heating at large 
densities, as originally emphasized by
Goldreich and Kwan (1974).  In this case it is easy to show that
$\gamma$ should be 
significantly {\it larger} 
than unity at high densities.  For example, at $n=(10^4,10^5$)
cm$^{-3},\ \gamma=(1.4,1.3)$, 
assuming $c=1.2$ and $T_{gr}\gg T$, and including the uncertain density
dependence of the cooling function in eq.\ (\ref{gamma}).
These results are uncertain because a reliable treatment of the grain
temperature requires a radiative transfer calculation.  

4. Extremely large densities.  As pointed out by Lis \& Goldsmith (1990) and
others, the gas-grain coupling at very large densities is so strong
that the gas temperature 
will simply follow the grain temperature, while the latter is
controlled by the radiation 
field.  So $\gamma=1$ may obtain at very large densities when
embedded protostars are present, if the ambient radiation field is not
coupled to the local 
gas density. If positive feedback between density and star formation rate
exists, then the local ultraviolet energy density will scale similarly. The
resulting grain temperature is only a weak function of energy density,
scaling roughly with the 1/5 power.  For example, if the local star
formation rate per unit volume scaled with the square of the density,
$\partial \log T/\partial \log n$ would be 0.4 and $\gamma$ would be 1.4.
Accurate observational determinations of the gas temperature and
density in high-density clouds could resolve the issue.

A final remark is in order. It is well known that the
Coriolis force (e.g., Binney \& Tremaine 1987) and the magnetic field
(e.g.\ Shu, Adams \& Lizano 1987; Mouschovias 1987) may act against
gravitational collapse. In fact, \VS\ et 
al.\ (1996) noted that inclusion of these effects restores a
``higher-$\gamma$'' behavior to the numerical simulations, when
considering the ability of the flow to collapse
gravitationally. However, the fact that in the present paper the
density pdfs of simulations including both these effects still exhibit
power-law tails on their high-density sides suggests that the local
production of large turbulent density
fluctuations generated by the stellar energy input is not strongly
inhibited by these processes. The effective compressibility of the
flow probably still has an equivalent polytropic exponent significantly 
smaller than unity.

\section{Applications}\label{applications}

\subsection{Can the density pdf be used to derive the stellar IMF?}

Padoan, Nordlund, and Jones (1997, PNJ) have attempted to derive the mass 
spectrum of collapsing gas clouds from the density pdf of simulations.  
The procedure is as follows.  If $f(\rho)$ is the density pdf, then 
$g(\rho)\sim\rho f(\rho$) is the  fraction of mass at a
given density.  Assuming that objects collapse if their mass exceeds the 
Jeans mass $M_J$, proportional to $\rho^{-1/2}$, they propose that the 
fraction of collapsing structures of mass $<M$ is proportional to
$\int^\infty_{\rho_J}g(\rho)d\rho$.  The IMF is then the distribution  of Jeans
masses, obtained by differentiating this integral, giving an IMF proportional
to $g(\rho)d\rho/dM_J$.  Since $d\rho/dM_J\propto M^{-3}_J$, this gives 
an IMF which is an $M^{-3}$ power law multiplied by the density pdf
expressed in  terms of Jeans mass, which they find from numerical
simulations to be a lognormal. This latter
factor is Gaussian in $\ln M^2$, with a mean that 
depends on the temperature and turbulent velocity dispersion.  This
dependence causes the logarithmic slope of the IMF to depend on these
parameters.
In principle this procedure could be applied to any density pdf, such as those 
discussed above. Our concern here  is the validity of
using a one-point density pdf to derive a distribution of masses, since mass 
as an attribute implies a coherent, contiguous ``object" or structure, while 
the density pdf contains no spatial information of this nature.  The
problem is basically that 
not just any region of a given density can form a collapsing object,
it must have a {\it size} 
large enough to contain a mass corresponding to at least the Jeans
mass at that density. 

The problem can be illustrated using a specific example.  Since
$M_J\propto\rho^{-1/2}$,  
the largest mass objects should form where the density is smallest.  The 
lowest density regions are the ``valleys" or ``voids" in the density field.  
Imagine the density field  as an array of pixels.  If we 
consider a density value far into the  low-density tail of the
density pdf, a lognormal form of the pdf indicates that only a small
fraction of  
pixels can have such small densities, and there is no {\it a priori}
reason to think  
that these pixels  will be sufficiently contiguous that there really exists 
a coherent region of that  density large enough
to contain the corresponding Jeans mass (or any specified mass).  In other 
words, in order to collapse, these low-density pixels must be spatially 
connected such that they form a very large
low-density region.  In general we expect the low density valleys to
be scattered  
through the density field in some way.  A Jeans' mass worth of matter
centered at the  
position of a low-density pixel would be expected to contain pixels of
a range of densities,  
some very large.  This inconsistency occurs at any density--the low
density case is  
just a severe example.  Even at pixels where the density is largest,
so the Jeans mass  
will be smallest, there is no reason to assume that nearby pixels out to mass 
$M_J$  will be at that density.\footnote{
  A similar problem has been pointed out about \VS's (1994) criterion for
  the development of hierarchical structure based on the density pdf
  (A.\ Noullez, private communication). Nevertheless, in that work, an
  additional constraint was required from the density pdf, namely, that
  it exhibited self-similarity as smaller and smaller regions are
  considered. This requirement may possibly provide 
  the necessary spatial information.} These same remarks apply to other
instability criteria (since 
it is not at all clear that the thermal Jeans mass plays any role;
e.g. Wiseman \& Adams 1994; Simon 1997; Chappell \& Scalo 1997).

PNJ apparently recognized this problem, and state that 99\% of the
density {\it peaks} in their simulations contain mass exceeding the
Jeans mass. However, our concern here is that, according to the
formulation used by PNJ, the high-mass portion of the IMF, which is
the region of most interest, is determined by {\it low}-density side
of the density pdf. The problem then is that the low-density part of the
pdf corresponds to {\it voids} in the density distribution, not peaks, and
these low-density regions would have to be 
coherent over extremely large scales to contain a Jeans mass (which is
largest at smallest densities). Furthermore, it is well-established
observationally that stars form in the {\it peaks} of the density
distribution (i.e., the clouds), not the voids. The implication that
the highest-mass stars form in the lowest-density regions contradicts
a large body of evidence on local star-forming complexes showing that
high-mass stars form preferentially in 
the {\it densest} regions (e.g. compare Orion and Taurus, or the
spatial distribution stars by luminosity within individual clusters in
Orion and other regions that contain massive stars).

Yet another concern is the statement itself by PNJ that the vast
majority of their density peaks is Jeans-unstable, since it has been
suggested on both observational (e.g., Falgarone, Puget \&  P\'erault
1992; Magnani, LaRosa \& Shore 1993) and numerical (\VS\ et al.\ 1997) grounds
that a large fraction of the turbulence-induced transient density
fluctuations are not gravitationally bound. Besides, PNJ's simulations
are not reported to be self-gravitating, so the statement that the
density peaks are unstable appears inconsistent.

We conclude that the one-point density pdf cannot be used to
derive a mass  function of ``clouds" or protostars: spatial statistics
are necessary.  One  
might consider the two-point density pdf, which gives the probability
that pairs of points  
separated by r have densities $\rho_1$ and $\rho_2$.  A second-order
moment of this two-point  
probability distribution, the correlation function, reduces this
information greatly (in the 
same way that  the variance reduces the one-point pdf to a single number).  
But even the 2-point pdf is  insufficient to estimate the mass
spectrum because we need to know 
the probability that there  exists a region with enough contiguous
pixels in a given density 
range to comprise a mass  sufficient for
instability.  At the least, a joint probability distribution of densities and 
coherence sizes would be needed, but construction of this joint
distribution is equivalent  
to directly constructing the mass spectrum from the simulations or
observational data.   
The plausibility of our conclusion can also be seen by considering the
process of estimating a  
mass spectrum for interstellar ``clouds" from observational data.
Within an arbitrary density  
field (or a three dimensional field with radial velocity as one axis,
as for spectral line  
data) the identification of entities defined in some operational
manner and the  
measurement of their attributes is a complicated and subjective
procedure (e.g. Houlahan and  
Scalo 1992, Williams et al. 1994), while the estimation of the column
density pdf by histogram  
construction for the same field is simple by comparison.  This
difference in computational 
effort is a reflection of the essential role of spatial information in
the former  procedure and the  absence of such information in the latter. 

        It is interesting to note that the procedure proposed to derive
the mass spectrum of bound condensations by PNJ is very similar in
approach to the 
derivation of the mass spectrum of bound objects in a gaussian density
field (in the linear regime) used in considerations of galaxy formation
(e.g. Press \& Schecter 1974, Peacock \& Heavens 1990).  There the
threshold density is taken as the critical density contrast needed at some
initial time so that the contrast will be about unity at a later time.  But
in the cosmological case, the derivation assumes that the density field is
succesively smoothed by window functions of scale R, which would correspond
to a mass M proportional to the mean density in the window.  This smoothing
takes care of the problem discussed here, of having enough extent at each
critical density to contain mass M, but the answer, even for a gaussian
random field, depends on the power spectrum and the rather arbitrary choice
of window function.  In the case of simulations the problem is more
difficult.  In particular, we suspect that because the density field is
non-gaussian, the answer will depend on higher-point correlations, as we
speculated earlier.
 
In any case, using simulations alone, we think that the only  way to determine 
the mass spectrum of coherent condensations, whether collapsing under the  
Jeans criterion or not, is to directly identify ``clouds" in the  simulation 
and calculate their frequency distribution (e.g. \VS\ et al.\ 1997;
Chappell and Scalo 1997).   There is no ``shortcut" to  
the IMF through the one-point density pdf.

Besides the conceptual problem of deriving an IMF from the one-point 
density pdf without spatial information, the IMF derived by PNJ does not 
give a good representation of the observed IMF in either field stars,
open clusters,  
OB associations, or nearby galaxies.  The PNJ IMF decreases with
increasing mass with a power  
law index that decreases (becomes more negative) with increasing mass.
PNJ recognized  
that the function does
not match the observed IMF at any single gas temperature and velocity 
dispersion, and so attempted to remedy the situation by assuming a   
T$^{-1}$ distribution of 
gas temperatures between 5 and 40 K and showed that such a temperature 
distribution could approximately match the Miller \& Scalo (1979)
field star IMF. The problem is that the Miller-Scalo  IMF at higher masses
has long been superseded by a large body of observations (see Scalo 1986 for 
a summary of work before 1985; also Rana 1991 for field stars, Massey 
et al. 1995a, b for massive stars in the
Galaxy and Magellanic Clouds; see Scalo 1997 for a recent review) that
suggest that for masses above about  
1.5 \m, the IMF is not lognormal, but is best fit by a power law of slope 
around $-1.7\pm 0.2$ with  (probably) a flatter
high-mass tail.  Such a form would be difficult to produce from the PNJ 
formalism without a
very contrived mixture of gas temperatures and velocity
distributions.\footnote{For comparison, if the density pdf were a
  power law of the form $f(\rho)\sim \rho^{-\theta}$,
  with a sharp cutoff at small densities, the IMF for a single-temperature
  Jeans mass would be proportional to $M^{2\theta-5}$, or $M^{-1.6}$ for
  $\theta=1.7$ as found here, in good agreement with observations.
  The flattening at large 
  masses could be provided by the flattening of $f(\rho)$ at densities near the
  peak.  The turnover of the IMF at small masses could be ascribed to
  ``turbulent viscosity'' steepening the density pdf at large densities (which
  requires $\theta>2.5$ in this regime if it were a power law).  By assuming a
  parameterized distribution of temperatures, we could match essentially any
  desired form of the IMF.  However we emphasize that this procedure is
  unwarranted, and only give the example as an illustration of the ease with
  which such a parameterized IMF model could produce a match to any set of
  observations.} 

For the above reasons, we must conclude that the application of the PNJ IMF 
to globular clusters (Padoan, Jimenez, and Jones 1997), low surface
brightness galaxies (Padoan, Jimenez, and 
Antonuccio-Delogu 1997), and other galaxy evolution problems (Chiosi et al.
1997) are without basis.  The galaxy evolution models of Chiosi et
al. (1997) do represent a major 
advance in such modelling, in that the coupling between the IMF and the
energy equation is explicitly treated.  However we interpret their results
as showing that an IMF that favors higher mass stars for some combination
of lower mean gas density and higher temperature and/or velocity dispersion
can account for a number of observed chemical and photometric constraints
for elliptical galaxies.  We only point out that there is as yet no
physically consistent model for such an IMF and that other models besides
the PNJ model could be constructed to give similar dependences. 

We emphasize that the fact that a theoretical 
formulation of a problem has a large enough number of adjustable parameters
to  interpret a number of diverse phenomena is not a measure of its
physical validity. A theoretical IMF that can be applied to these and other 
problems must be able to give a natural  account of contemporary
studies of the local field star and cluster IMF, and, besides, cannot 
be derived solely from the density pdf. A brief summary of recent
theoretical models for the IMF is given in Scalo (1997).

\subsection{Applications to Column Density Statistics}
\begin{figure}[htb]
\plotone{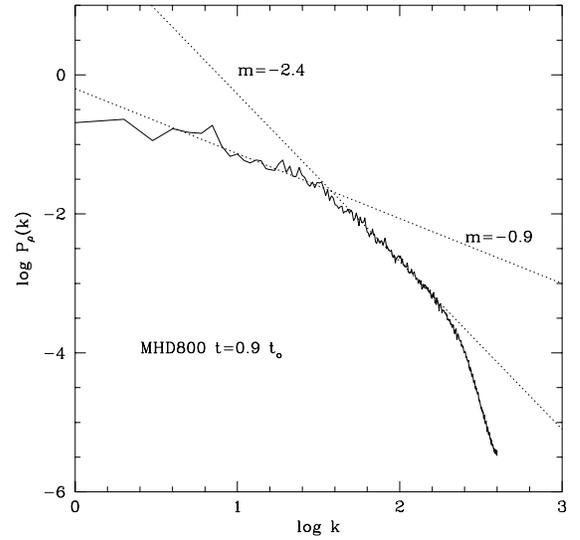}
\caption{Density power spectrum for run MHD800 at time $t=0.9
  \to$. Two power laws are seen, one in the range $0.6 \leq \log k
  1.5$ ($k$ is the wavenumber) with slope $\sim -0.9$, and the other
  in the range $1.5 \leq \log k \leq 2.3$, with slope $-2.4$. The
  region with the steeper slope may actually not be a true power law,
  but rather a slow exponential decay due to the mass diffusion term in the
  continuity equation.} 
\label{dens_spec}
\end{figure}

        Our critique of the PNJ IMF does not extend to their interpretation 
of the increase of extinction standard deviation with average extinction 
found by Lada et al.
(1994), given by Padoan, Jones, \& Nordlund (1997).  We agree that this 
increase constrains the
density pdf and probably requires an intermittent tail (not necessarily 
lognormal, though) and
a power spectrum of the density field $P(k)\propto k^x$, with $x$ around 
$-2$ to $-3$.  However, whether or not simulations with different physical 
assumptions can give the requisite
intermittency and power spectrum remains an open question.  We have 
examined the power spectrum
of a 2-dimensional  MHD model (fig.\ \ref{dens_spec})
and find a small intermediate range of wavenumbers
over which the power spectrum is a power law with index around $-2.4$, 
close to the value of $-2.6$ reported by Padoan,
Jones, and Nordlund (1997) in their simulations, although a much flatter 
and just as extensive
power law occurs at smaller wavenumbers in the present
simulations. A more direct test of the
simulations would be an estimate of the pdf of column densities.  Such a 
comparison, using IRAS 100 $\mu$m column densities for low-mass star regions 
and $^{13}$CO data for  high-mass star regions,
is postponed to a separate paper (Scalo, Chappell, Miesch, and 
\VS, in preparation). 

Note that in our simulations, the wavenumber range with a $-2.6$
``slope'' may actually constitute the range at which the mass
diffusion is active, producing an extended exponential decay which may
be confused with a power-law. This effect could also be at work in the
simulations of Padoan, Jones, \& Nordlund(1997), due to the numerical
diffusion. Note, however, that this cannot be at
the origin of the power-law tails we observe in the density pdfs from
our simulations, since the effect of the diffusion term is to {\it
  smooth out} fluctuations, while the power-law tails in the pdfs are
{\it enhancements} over the lognormal alternative.
 
\subsection{Size distribution of wind-driven shells}
Shells driven by H II regions, protostellar winds, supernova remnants
(SNRs), and superbubbles 
are believed to play an important role in shaping interstellar
structure over a range of size 
scales, and the velocity and size distributions of such shells are of
interest in a variety of 
astrophysical contexts, such as the porosity of the ISM (Oey \& Clarke
1997) and models for 
the IMF (Silk 1995).  The expansion law for shells depends on the
ambient interstellar 
density, so a distribution of ambient interstellar densities could
affect the predicted size and 
velocity distributions.  The distribution of sizes of SNRs and
superbubbles for a distribution 
of source luminosities has been treated in detail by Oey \& Clarke
(1997), who assumed a 
constant ambient density.  Here we illustrate the role of a density
pdf by assuming a constant 
luminosity. For simplicity, we neglect the effect of shell stalling
which occurs when the 
driving pressure matches the ambient pressure.  
We are
mostly interested in effects that occur in molecular clouds subjected
to internal protostellar 
winds.  In this case the appropriate expansion law is probably that
for a non-adiabatic 
momentum-conserving shell (see the conditions derived in Norman \&
Silk 1980), which can be expressed as
\begin{equation}
R\sim L^{1/4}\rho^{-1/4}t^{1/2},
\end{equation}
\ni where $L$ is the source luminosity, which we assume is constant
for illustrative purposes, and $R$ the radius of the shells. 

For a constant star formation rate $B$, the size distribution for
a given ambient density is given by $N(R,\rho)dR=Bdt$, so, 
integrating over the pdf of ambient densities,
\begin{equation}
N(R)\sim\int^{\rho_2}_{\rho_1}f(\rho)(dR/dt)^{-1}d\rho.
\end{equation}

If the ambient densities were the same for all sources ($f(\rho$) a
delta function), then, 
since eq.\ (15) gives ($\partial R/\partial t)^{-1}\sim R$, $N(R)$
would scale as $R$, reflecting 
the larger number of shells at small velocities.  For the case of a
distribution of ambient 
densities, the integration limits require some care.  We assume that
the lower and upper 
limits of ambient densities are $\rho_{min}$ and $\rho_{max}$, and
that all the sources are 
only active for a time $t_e$.  Then for radii $R$ less than some small
$R=R_1$, all ambient 
densities, even up to the maximum density, allow expansion to radius
$R$ in a time $t_e$.  But 
for larger radii, shells expanding into high density material will not
reach that size in time 
$t_e$, so the limits of integration are $\rho_{min}$ to $\rho(R,\
t_e$), where $\rho(R,\ t_e$) 
is the ambient density for which a shell will reach size $R$ within
time $t_e$.  So 
\begin{equation}
N(R)\sim\int^{\rho(R,\ t_e)}_{\rho_{min}}f(\rho)\rho^{1/2}R d\rho.
\end{equation}
\ni Taking $f(\rho)\sim\rho^{-\theta}$ gives
\begin{equation}
N(R)\sim\left[\rho(R,t_e)^{-\theta+3/2}-\rho_{min}^{-\theta+3/2}\right]R.
\end{equation}
\ni For $\theta>3/2$ (steeply declining pdf), most of the ambient
material is at small $\rho$, 
and the second term dominates.  The size distribution is then
unaffected by the density pdf, $N(R)\sim R$, because the pdf is in
effect a delta function. 

But if $\theta<3/2$ (flatter pdf), the first term dominates.  Since
$\rho(R,t_e)\sim R^{-4}$ 
for a fixed $t_e$, the resulting size distribution is
\begin{equation}
N(R)\sim R^{4\theta-5}.
\end{equation}
\ni For $\theta<3/2,\ 4\theta-5<1$, so $N(R)$ increases less rapidly
than $R$; shells at small 
$R$ become significant because more of the ambient material is at
large densities.  The same 
derivation for an adiabatic pressure-driven shell (Weaver et al.\ 1977)
gives $N(R)\sim R^{5\theta-8/3}$ for $\theta>2/3$.

The situation is actually much more complicated for a number of
reasons.  The density pdfs 
found in the simulations are strongly decreasing functions of $\rho$
for large densities, but 
increasing, and extremely uncertain, functions of $\rho$ at small
densities.  Numerical 
integration of eq.\ (17) for a double power-law or lognormal density
pdf would be a useful 
exercise, but our purpose here is only to illustrate the potential
effect.  Secondly, a 
distribution of source luminosities should be included (i.e. a
generalization of Oey \& 
Clarke's 1997 result to include the density pdf).  Third, as noted
above, shell interactions 
are probably important (see also Norman \& Silk 1980), and for this
process the effect of a 
distribution of ambient densities would be to affect the column
density and velocity 
distributions of the interacting shells.  Finally, we note that the
present derivation suffers 
from the same lack of spatial information that we claimed precludes a
derivation of the mass 
spectrum.  If the ambient density fluctuations were all on scales
smaller than any shell size 
of interest, then the effect of the density distribution would only be
to corrugate the shell 
in some irregular manner, as different parts of a given shell expand
into different ambient 
densities.  On the other hand, some coherent density fluctuations in
the ISM have scales 
larger than the shell sizes of interest.  A full examination of these
problems is beyond the scope of the present paper.

\subsection{Comparison with cosmological simulations} \label{compar_cosmo}

It is of interest to compare our results with cosmological
simulations of the evolution of density fluctuation in the universe.  Such
calculations usually contain a minority of the matter in baryonic form,
with the rest either in ``cold" or ``hot" dissipationless particles meant to
represent varying dark matter candidates, or some combination of the two. 
The dissipationless component is equivalent to a self-gravitating Burgers
flow (no pressure, $\gamma = 0$) with viscosity arising only from numerical
diffusion, while the coupled baryonic component follows the Navier-Stokes
equation with self-gravity.

Most of the work in this field has been aimed 
at using the density pdf as a discriminator of parameters related to
initial conditions, such  
as the index of the initial power spectrum or deviations from initial
Gaussian statistics.   
Several papers, apparently beginning with Hamilton (1985), have
proposed that the density  
pdf in the not-too-strongly nonlinear regime is lognormal (see
Weinberg and Cole 1992,  
Kofman et al. 1994, Bernardeau and Kofman 1995; for other references
see Protogeros and  
Scherrer 1997).  However  Bernardeau and Kofman (1995) argued that it
is likely that the 
lognormal  distribution only obtains for an initial power spectrum
with index not too different  
from $-1$ (as in cold dark matter scenarios) and for a variance of the
(smoothed) density field 
not too  large (less than around unity).  For conditions outside of
this range the density pdf 
appears  to have a power-law, not lognormal, high-density tail
(Bouchet and Hernquist 1990, 
Melott et al.  1997; these power laws are to be distinguished from the
steeper power laws 
that result  from caustics in the Zeldovich approximation with no
smoothing). Although in general such simulations are not isothermal
and in some cases refer to pressureless dark matter,
the apparent transition from lognormal at small density 
variance to power law at larger density variance may be understood
as the analogue of the Mach number dependence described in the present
paper, the range of density values being just too small for the
power-law range to develop.

A recent discussion of a cold-plus-hot dark
matter simulation that presents the density pdf has been given by Klypin, 
Nolthenius, and Primack
(1997).  These simulations use a particle-mesh code with a force mesh
resolution of $512^3$ in three dimensions.  The resulting function
$\rho f(\rho)$
(their fig.\ 16) shows a clear power-law over more than two orders of
magnitude in density, with a power-law index of about $-1.2$, and a dropoff
at the highest densities due to finite resolution.  There is certainly no
hint of a lognormal form.  The result supports our contention of a power
law density pdf driven by advection, and shows that power-law behavior
occurs even in three dimensions.  The power law index is steeper than found
here by about 0.2 to 0.6, but it is uncertain whether this is a result of the
different dimensionality, the fact that the self-gravity is mainly supplied
by the dissipationless component, or some other effect.

\section{Conclusions}\label{conclusions}

In this paper we have discussed various theoretical aspects of the density
probability distribution function, or pdf, of the ISM. We first
presented evidence that the pdf appears to be a power-law  at
high densities in a number of two-dimensional numerical simulations of
the ISM at intermediate scales, in which the various intervening
physical agents are progressively removed until only the nonlinear
advection operator of the momentum equation is left. Additionally, in
all cases the morphology is extremely filamentary. This result has
two main implications. First, it suggests that both the functional form
of the pdf and the morphology of the density field are the signature
of the advection operator. 

Second, this result is in contrast with previous findings that the pdf
is lognormal (\VS\ 1994; Padoan et al 1997 a, b, c). To resolve the
discrepancy, we turned to the one-dimensional simulations of PVS, 
which suggest that the lognormal pdf is realized only in
the isothermal case ($\gamma=1$), while at smaller values of $\gamma$
a power law appears. Moreover, PVS report that the deviation from a
lognormal is larger at larger Mach numbers. These results are
consistent with the fact that the simulations of \VS\ (1994) and Padoan 
et al.\ (1997a, b, c) were isothermal, while those presented here in
all cases have small or zero values of $\gamma$. 

The dependence on the polytropic index $\gamma$ prompted an
investigation on what are its expected values in the actual ISM, as
determined by the equilibrium between heating and cooling rates, which 
in turn depends on the density and temperature of the medium. From
inspection of 
published heating and cooling rates, we expect that at densities below
$\sim 10^3$ cm$^{-3}$  
values of $\gamma$ significantly smaller than unity should occur,
while at larger densities, 
values near unity should appear. The switch-over occurs because of the
effect of radiative 
trapping in  decreasing the density dependence of the molecular
cooling rate at densities above  
$\sim10^3$ cm$^{-3}$. At densities larger than around $10^4$
cm$^{-3}$ gas-grain 
collisional  cooling or heating should dominate, forcing $\gamma$ back away 
from unity.  Thus, lognormal pdfs are only expected in the range $10^3$ to 
$(1-5)\times10^4$ cm$^{-3}$, with power  law pdfs outside of this
range of densities. However, at extremely
large densities the gas-grain coupling may be so efficient that the gas
temperature will simply follow the grain temperature which is controlled by
the ambient UV radiation field.  To the extent that the radiation field
energy density is independent of the gas density, the temperature will then
be constant, so a return to a lognormal density pdf may occur at such large
densities.  A star formation rate-density coupling would increase $\gamma$
somewhat above unity through the effect of the young massive stars on the
grain temperature.

A number of possible applications were discussed. Most
importantly, we discussed the feasibility of deriving the mass
spectrum of bound condensations or the stellar IMF
from knowledge of the density pdf alone, as recently done by PNJ. We
conclude that this is not 
possible, because the pdf does not contain spatial information, while
the gravitational instability criterion requires simultaneous
knowledge of mass and size information. Second, we briefly discussed the
applicability to extinction statistics (Padoan, Jones \& Nordlund
1997).  Third, we analyzed the effect of an inhomogenous density field on
the size distribution of wind-driven shells in the ISM. We find that
in a simple example case with $f(\rho)\sim \rho^{-\theta}$, the
resulting distribution increases more 
slowly with shell size than in the case of a uniform density
background if $\theta < 3/2$. However, we warn that the real situation
is likely to be much more complicated.

Finally, we compared our results with those from a number of
cosmological simulations, in which both lognormal and power-law forms
of the pdf have been reported. We speculate that the appearance of
lognormal pdfs for non-isothermal situations may be a consequence of
small density variances, analogous to the Mach number effect
described here.

        A central result of our work is the conclusion that it is primarily
the  nonlinear advection operator operating on a source of compressible
motions (whether it be self-gravity or nonlinear waves or cooling or
something else) that is responsible for both the general filamentary
morphology (which would include sheets in 3D) and for the power law portion
of the density pdf.  This result is consistent with the presumed tendency
of the advection term to induce multiplicative processes which lead to
power law scaling, as it does in the incompressible case. If this is
correct, then the simulations studied here should lead to investigations of
analytical models for the behavior of the advection operator in a highly
compressible medium.  For that purpose our simulations provide a constraint
in the form of the density pdf that must be explained both when $\gamma$ is
and is not close to unity, as well as a constraint on the morphology. As an
example, we point out that Elmegreen's (1997) construction of a fractal by
nesting can be shown to produce a density pdf that has the form
$f(\rho)\sim \rho^{-2}$, independent of fractal dimension, which is not much
different from the power laws found here. Although Elmegreen's model is a
static construction with no reference to underlying physics such as
advection, this does suggest that an advection model that is related to
hierarchical nesting can explain the power-law density pdf.

Alternatively, the physics might be multiplicative but not involve
nesting. For example, the action of advection could simply be to steepen
fluctuations into ``shells'', or ``filaments'' or ``spikes'' (depending on
dimensionality), as occurs in the Burgers equation.  After one crossing
time the evolution is dominated by the merging of these entities (which we
might refer to as ``blobs''), a process that is described by the so-called
``coalescence equation''.  From much previous work we know that the solution
of this kinetic equation, when forced at small scales, is a power law mass
spectrum with index around $-1.5$ to $-2$. If the blobs were spikes or
filaments or shells whose thicknesses were relatively constant or at least
independent of their mass, the density pdf would be a power law in the same
range. Obvious variations and elaborations on this theme suggest
themselves.

Our point is that it is relatively easy to think of simple models
that might explain a power law density pdf.  However much more work is
needed to understand the actual physics of advection acting on a source of
compressibility, and how that physics gives rise to the observed and
simulated morphology and statistical properties.  Given the longstanding
difficulties with the analogous questions in the incompressible case (in
which vorticity dynamics plays the key role), we do not expect
oversimplified models to provide the necessary answers.

\begin{acknowledgements}
We thank Paul Shapiro and Bruce Elmegreen for useful comments. The HD
and MHD runs were performed on the CRAY YM-P 4/64 of DGSCA, UNAM.
This work was supported by NASA Grant NAG5-3107 and a grant from Cray Research
to J.\ S.\ , grants UNAM-DGAPA IN105295 and UNAM/CRAY SC007196 to
E.\ V.-S , a joint CNRS-CONACYT grant to E.\ V.-S and T.\ P.,  and
by a grant from the  PCMI CNRS program to T.\ P.
\end{acknowledgements}

\vfill\eject

\vfill
\eject

\vfill
\eject

\vfill
\eject

\vfill
\eject

\end{document}